\documentclass[a4paper,11pt]{article}
\usepackage{silence}
\usepackage{jheppub} 
\usepackage{subcaption} 
\usepackage{booktabs}
\usepackage[acronym]{glossaries}
\setkeys{glslink}{hyper=false}
\usepackage{cleveref}
\usepackage{placeins}

\newcommand{\pt}{\ensuremath{p_{\mathrm{T}}}}
\newcommand{\myy}{$m_{\gamma\gamma}$}

\newcommand{\ifb}{\ensuremath{\mathrm{fb}^{-1}}}

\newacronym{AD}{AD}{anomaly detection}
\newacronym{CATHODE}{CATHODE}{Classifying Anomalies THrough Outer Density Estimation}
\newacronym{SIC}{SIC}{Significance Improvement Characteristic}
\newacronym{MADE}{MADE}{Masked Autoencoder for Distribution Estimation}
\newacronym{AE}{AE}{autoencoder}
\newacronym{VAE}{VAE}{Variational Autoencoder}
\newacronym{MLP}{MLP}{Multilayer Perceptron}
\newacronym{KL}{KL}{Kullback-Leibler divergence}
\newacronym{BDT}{BDT}{Boosted Decision Tree}
\newacronym{SR}{SR}{Signal Region}
\newacronym{SB}{SB}{Sideband Region}
\newacronym{SM}{SM}{Standard Model}
\newacronym{BSM}{BSM}{Beyond the Standard Model}
\newacronym{MC}{MC}{Monte Carlo}

\newcommand{\nonresyyjjj}{\ensuremath{\gamma\gamma+\mathrm{jets}}}            
\newcommand{\nonresttyy}{\ensuremath{\gamma\gamma\,t\bar{t}}}                            
\newcommand{\nonresllyyjj}{\ensuremath{\gamma\gamma\,\ell\ell}} 
 
\newcommand{\ttH}{\ensuremath{t\bar{t}H}}

\newcommand{\WNHyyNall}{\ensuremath{\tilde\chi_{1}^\pm(W\tilde\chi_{1}^0)\,\tilde\chi_{2}^0(H\tilde\chi_1^0)}}
\newcommand{\WNHyyN}[1]{\ensuremath{\tilde\chi_{1,#1}^\pm(W\tilde\chi_{1}^0)\,\tilde\chi_{2,#1}^0(H\tilde\chi_1^0)}}
 
\newcommand{\XSHall}{\ensuremath{X\to S(b\bar{b})H}}

\newcommand{\XSHllall}{\ensuremath{X\to S(qq)H}}

\newcommand{\HlHyylall}{\ensuremath{\tilde\chi_{1}^{+}(H\ell^{+})\,\tilde\chi_{1}^{-}(H\ell^{-})}}

\newcommand{\WlZvHvHyylall}{\ensuremath{\tilde\chi_{1}^\pm(H\ell^\pm)\,\tilde\chi_{1}^0(W\ell/Z\nu)}}
\newcommand{\WlZvHvHyyl}[1]{\ensuremath{\tilde\chi_{1,#1}^\pm(H\ell^\pm)\,\tilde\chi_{1,#1}^0(W\ell/Z\nu)}}
 
\newcommand{\ttFCNCall}{\ensuremath{t\bar{t},\ t\to qH\ (q=u,c)}}

\newcommand{\thFCNCall}{\ensuremath{pp\to t(bW)H}}

\newcommand{\TTtZNtHyyNall}{\ensuremath{\tilde t\tilde t\to t\tilde\chi_{2}^0(Z\tilde\chi_{1}^0)\,t\tilde\chi_{2}^0(H\tilde\chi_{1}^0)}}

\newcommand{\XHHall}{\ensuremath{X\to H(b\bar{b})H}}

\newcommand{\BBbZNbHyyNall}{\ensuremath{\tilde b\tilde b\to b\tilde\chi_{2}^0(Z\tilde\chi_{1}^0)\,b\tilde\chi_{2}^0(H\tilde\chi_{1}^0)}}

\newcommand{\CCcZNcHyyNall}{\ensuremath{\tilde c\tilde c\to c\tilde\chi_{2}^0(Z\tilde\chi_{1}^0)\,c\tilde\chi_{2}^0(H\tilde\chi_{1}^0)}}

\newcommand{\HVTVcXjjHyyall}{\ensuremath{V'^{\pm}\to H\,X(q\bar{q})}}
\newcommand{\HVTVcXjjHyy}[2]{\ensuremath{V'^{\pm}_{#1}\to H\,X_{#2}(q\bar{q})}}
 
\title{Towards anomaly detection searches for new physics signatures including Higgs bosons with weakly supervised machine learning}

\author[a,b]{Chi Lung Cheng,}
\author[a]{Julia Gonski,}
\author[c]{Julie Khalilieh Romman,}
\author[d]{Runze Li,}
\author[a]{Qibin Liu,}
\author[a,b]{Benjamin Nachman,}
\author[a,b]{Dennis Noll,}
\author[a,e]{and Liangyu Wu}
\affiliation[a]{Fundamental Physics Directorate, SLAC National Laboratory, Menlo Park, CA 94025, USA}
\affiliation[b]{Department of Particle Physics and Astrophysics, Stanford University, Stanford, CA 94305, USA}
\affiliation[c]{San Jose State University, 1 Washington Square, San Jose, CA 95192, USA}
\affiliation[d]{Department of Physics, Yale University, New Haven, CT 06511, USA}
\affiliation[e]{Department of Physics, Stanford University, Stanford, CA 94305, USA}
\abstract{
The Higgs boson, with its universal coupling to mass, provides a broadly applicable portal to sectors beyond the Standard Model and is therefore a natural anchor for anomaly detection (AD) at collider experiments.
The Higgs And X Anomaly Detection (HAXAD) strategy offers a principled approach to searching for such anomalies occurring in association with a Higgs boson by combining machine-learning-based feature embedding, background estimation, and weakly supervised classification.
This work extends the previous HAXAD approach towards the level of maturity required for application to recorded collider data. A major addition is the introduction and comparison of two new embedding strategies, which in turn shape the background estimation and classification.
In addition, a new inference framework is developed, yielding signal-agnostic and signal-specific cross section limits and thereby completing the statistical machinery needed for future AD analyses built on HAXAD.
The set of investigated signal models is also significantly expanded, allowing for the evaluation of sensitivity on a much broader phase space.
Improvements to the method increase signal sensitivity with respect to the original method, and when benchmarked against an example cut-based search on the same final state, HAXAD matches or exceeds the best individual cut-based limits for a wide variety of considered signal models.
These developments strengthen the case for HAXAD as a viable and compelling AD-based search strategy with novel discovery potential at colliders.
}

\begin{document}
\maketitle
\flushbottom

\section{Introduction}
\label{sec:intro}

The discovery of new \gls{BSM} physics is a key objective of research at high energy particle colliders. 
Search programs at modern collider experiments comprise signal model-focused analyses targeting key physics drivers such as the nature of dark matter~\cite{Boveia:2018yeb} or the hierarchy problem~\cite{Craig:2023lyt}. 
Increasingly more generic anomaly detection (AD)-based approaches are also incorporated, which seek to identify deviations within a known background distribution without relying on particular signal features. 
Machine learning (ML) offers a robust means to develop and implement AD techniques, with broad utility across high energy physics~\cite{Belis_2024,karagiorgi2021machinelearningsearchnew,Kasieczka_2021,Aarrestad_2022}. 

Higgs boson final states provide a particularly generic and thus compelling handle for AD at colliders. 
As the only fundamental scalar and a field whose couplings set the masses of \gls{SM} particles, the Higgs boson naturally acts as a portal~\cite{Patt:2006fw} through which many BSM sectors can communicate with the SM. 
Moreover, several Higgs decay modes such as $H\to\gamma\gamma$ offer clean reconstruction and good mass resolution, enabling precise comparisons between data and reference distributions and thereby amplifying the reach of AD methods that search for localized or subtle discrepancies in high-dimensional feature space.

To achieve agnosticism to signal models, the implementation of ML-based AD requires an approach to training that does not rely on full and correct labels on training instances. 
Unsupervised ML trains directly on data without any labeled instances.
Commonly implemented with autoencoders~\cite{Farina:2018fyg,Heimel:2018mkt} which aim to reconstruct input instances through a lower-dimensional latent space, unsupervised methods provide the most generic approach to selecting anomalies based only on background incompatibility, but suffer from limited interpretability and potential sensitivity to modeling effects. 
Weakly supervised methods such as the classification without labels (CWoLa)~\cite{Metodiev:2017vrx,Collins_2018,Collins_2019} represent a middle ground wherein signal models are used in classifier training, but in datasets mixed with background in such a way that labels are applied but noisy.
Finally, semi-supervised training includes the partial use of correctly labeled signal instances~\cite{Kuusela:2011aa,Park:2020pak}. 
These categories of AD have been implemented in searches by the ATLAS~\cite{PhysRevD.108.052009,2yq5-vj59,PhysRevLett.132.081801} and CMS~\cite{cmsad_2025} experiments.

This work leverages the Higgs And X Anomaly Detection (HAXAD)~\cite{cheng2025weaklysupervisedanomalydetection} analysis strategy to target anomalous events which include an SM Higgs boson that decays to two final-state photons. 
This event category was previously targeted in ATLAS with the Run~2 dataset, using a variety of non-orthogonal event selections based on standard collider observables to provide model-independent sensitivity~\cite{ATLAS:2023omk}.
HAXAD combines ML-based feature embedding, background estimation based on \gls{CATHODE}~\cite{Hallin:2021wme} using the Higgs boson mass sidebands, and the weakly supervised classifier approach of CWoLa for the enrichment of a region with anomalous events.

With respect to the initial HAXAD study, this work expands in a variety of directions that increase sensitivity and validity for use in a search with collider datasets. 
Additional inputs are used to train the first-stage feature embedding model, specifically providing information about leptons and heavy-flavor jets which expands the signal model profile to which HAXAD is sensitive. 
The embedding model is improved, and two new approaches are introduced with different levels of training supervision, offering complementary benefits that can be customized to the needs of a given analysis.
As the performance of an AD method is predicated on its breadth of sensitivity, a wider variety of signal models are generated to further qualify HAXAD performance.
Finally, upper limits on signal model cross sections are set and put in the context of the Run~2 ATLAS search targeting the same phase space. 
The final limits target an integrated luminosity of 470\,\ifb, projecting the impact of a HAXAD-based search considering a dataset roughly corresponding to that of the Large Hadron Collider (LHC) multipurpose detectors over Runs~2 and Run~3. 
\section{Samples}
\label{subsec:samples}

Monte Carlo (MC) simulated samples of proton--proton collisions at $\sqrt{s} = 13\,\mathrm{TeV}$ are used throughout this study.
The dominant SM background in the diphoton final state is non-resonant QCD $\gamma\gamma$ production (\nonresyyjjj).
It is generated with \textsc{MadGraph5\_aMC@NLO} in version 3.5.15~\cite{Alwall:2014hca} at leading order (LO) in the matrix element with zero, one, two and three additional partons, and showered with \textsc{Pythia} 8.312~\cite{pythia83-1,pythia83-2}.
The MLM merging scheme~\cite{Mangano:2006rw,Alwall:2007fs} is used to remove the double counting between the matrix-element partons and the parton-shower emissions.
The inclusive cross section is normalized to the next-to-leading-order (NLO) prediction.
Detector effects are modeled with \textsc{Delphes}\,3.5.1~\cite{deFavereau:2013fsa} in a configuration corresponding to the ATLAS experiment, modified to reconstruct large-radius jets with the anti-$k_t$ algorithm~\cite{Cacciari:2008gp,Cacciari:2011ma} and radius parameter $R=1.0$ in addition to the standard small-radius ($R=0.4$) jets.
Additional QCD backgrounds with charged leptons, neutrinos or top quarks in the matrix element are generated to enhance the statistics in specific regions of phase space: the $\gamma\gamma\,\ell\ell$ process, where $\ell$ denotes an electron, a muon or the associated neutrino, is simulated at LO with up to two additional partons, and the $\gamma\gamma\,t\bar{t}$ process is simulated with no additional partons.
The resonant SM background originates from Higgs-boson production with $H\to\gamma\gamma$; the ggH, VBF, VH and ttH modes are simulated with \textsc{Pythia}\,8.312.
Other SM processes are neglected owing to their small contribution and because the aim of this study is to demonstrate the anomaly detection strategy rather than to deliver a precision measurement.
The SM processes are summarized in the upper block of Table~\ref{tab:samples}.

The pseudo-data are produced with an unweighting procedure to emulate the observed data.
The expected yield of each process is scaled by its cross section and the assumed integrated luminosity of 470\,\ifb.
Events are then sampled from the MC datasets, according to their simulation weight, to build the pseudo-data from the QCD $\gamma\gamma$ and SM Higgs processes.
In this study, we apply a hypothetical Run~2 + Run~3 luminosity of 470\,\ifb. A total number of $\sim 2.1$ million pseudo-data events is produced from unweighting.

A wide variety of BSM signal processes are simulated under the same conditions to benchmark the analysis; they are summarized in the lower block of \cref{tab:samples}.
The analysis is designed to be model-independent: the signals are chosen to span a broad range of production mechanisms, final states and kinematic regimes, so as to probe the sensitivity as generally as possible rather than to target any individual signal.
The BSM signal models are drawn from several classes of theory: extended Higgs sectors and scalar resonances (\XHHall\ and \XSHall, and \XSHllall\ in which the scalar $S$ only decays to light quarks)~\cite{Robens:2019kga,Basler:2018dac,Baum:2019pqc,Chacko:2005vw,Branco:2011iw}, heavy vector triplets (HVT)~\cite{Pappadopulo:2014qza}, Minimal Supersymmetric Standard Model (MSSM)~\cite{Martin:1997ns} and $R$-parity-violating (RPV)~\cite{Barbier:2004ez} supersymmetry (SUSY), colored SUSY~\cite{Alves:2011wf} with stop, sbottom or scharm pair production followed by a neutralino cascade, and top-quark flavor-changing neutral currents (FCNC)~\cite{Aguilar-Saavedra:2004mfd}.  
In every case at least one SM-like Higgs boson is produced and decays to two photons, forced at generator level. The branching ratio of $H \to \gamma\gamma$ is included in the cross section throughout the following study.
We emphasize that the analysis is not tuned for this particular set of signals but is intended to generalize to any final state containing such a diphoton resonance.

A common pre-selection, emulating the standard ATLAS trigger in the $H\to\gamma\gamma$ analysis and reconstruction requirements~\cite{ATLAS:2022tnm}, is applied to all samples.
Events must contain at least two photons with $p_T > 22\,\mathrm{GeV}$ and $|\eta| < 2.5$.
The two leading photons must satisfy a diphoton trigger emulation --- the leading (sub-leading) photon with $p_T > 35~(25)\,\mathrm{GeV}$ --- and pass the relative-$p_T$ requirements $p_T^{\gamma_1}/m_{\gamma\gamma} > 0.4$ and $p_T^{\gamma_2}/m_{\gamma\gamma} > 0.3$. 
The diphoton invariant mass is required to fall in the window $105 < m_{\gamma\gamma} < 160~\mathrm{GeV}$.
Jets are reconstructed with $p_T > 25~\mathrm{GeV}$ and $|\eta| < 4.4$, and light leptons (electrons and muons) with $p_T > 10~\mathrm{GeV}$.

\begin{table*}[htbp]
  \centering
  \caption{
    Standard Model background (upper block) and BSM signal benchmark (bottom block) processes.
    Masses are given in GeV.
    Leading-order (LO) and next-to-leading-order (NLO) denote the QCD accuracy. The notation ($0,1,2j@\mathrm{LO}$) indicates that matrix elements with zero, one, or two additional partons are calculated at LO in QCD. Parentheses indicate particle decay channels.
    The \nonresyyjjj\ cross section is normalized to its NLO prediction.
    \XHHall\ is a CP-even scalar; \XSHall\ assumes $S\to b\bar b$, whereas \XSHllall\ forces $S\to q\bar q$ decays with $q=u,d$.
    The \HVTVcXjjHyyall\ benchmark follows the Heavy vector triplet (HVT) Model A in the charged production mode.
  }
  \label{tab:samples}
  \footnotesize
  \begin{tabular}{@{}ll@{}}
    \toprule
    Process & Channels and Masses \\
    \midrule
    \multicolumn{2}{@{}l}{\emph{Standard Model background}}\\
    \nonresyyjjj& $\gamma\gamma + 0,1,2,3j@LO$ \\
    \nonresllyyjj & $\gamma\gamma\,\ell\ell + 0,1,2j@LO$, $\ell = e,\mu,\nu$ \\
    \nonresttyy & $\gamma\gamma\,t\bar{t} + 0j@LO$ \\
    SM Higgs & ggH, VBF, VH, \ttH \\
    \midrule
    \multicolumn{2}{@{}l}{\emph{Extended Higgs sectors and scalar resonances~\cite{Robens:2019kga,Basler:2018dac,Baum:2019pqc,Chacko:2005vw,Branco:2011iw}}}\\
    \XHHall & $m_X = 260,\,280,\,500,\,1000$ \\
    \XSHall & $(m_X,m_S) = (500,100)$ \\
    \XSHllall & $(m_X,m_S) = (750,100)$ \\
    \addlinespace
    \multicolumn{2}{@{}l}{\emph{Heavy vector triplet (HVT)~\cite{Pappadopulo:2014qza}}}\\
    \HVTVcXjjHyyall & $(m_{V'},m_X) = (2000,1700),\,(2000,300),\,(500,300),\,(500,10)$ \\
    \addlinespace
    \multicolumn{2}{@{}l}{\emph{MSSM SUSY~\cite{Martin:1997ns}}}\\
    \WNHyyNall & $m_{\tilde\chi_1^\pm}=m_{\tilde\chi_2^0}=150,\,200,\,300,\,600$; $m_{\tilde\chi_1^0}=0.5$ \\
    \addlinespace
    \multicolumn{2}{@{}l}{\emph{RPV SUSY~\cite{Barbier:2004ez}}}\\
    \HlHyylall & $m_{\tilde\chi_1^\pm} = 100,\,150,\,300,\,450$ \\
    \WlZvHvHyylall & $m_{\tilde\chi_1^\pm}=m_{\tilde\chi_1^0}=200,\,400,\,600$ \\
    \addlinespace
    \multicolumn{2}{@{}l}{\emph{Colored SUSY~\cite{Alves:2011wf}}}\\
    \TTtZNtHyyNall & $(m_{\tilde t},m_{\tilde\chi_2^0},m_{\tilde\chi_1^0}) = (500,180,50),\,(1000,205,60),\,(1200,205,60)$ \\
    \BBbZNbHyyNall & $(m_{\tilde b},m_{\tilde\chi_2^0},m_{\tilde\chi_1^0}) = (500,180,50),\,(1000,205,60),\,(1200,205,60)$ \\
    \CCcZNcHyyNall & $(m_{\tilde c},m_{\tilde\chi_2^0},m_{\tilde\chi_1^0}) = (500,180,50),\,(1000,205,60),\,(1200,205,60)$ \\
    \addlinespace
    \multicolumn{2}{@{}l}{\emph{Top-quark FCNC~\cite{Aguilar-Saavedra:2004mfd}}}\\
    \thFCNCall &  \\
    \ttFCNCall &  \\
    \bottomrule
  \end{tabular}
\end{table*}

\section{Analysis Strategy}
\label{sec:analysis}

The analysis follows the HAXAD strategy first introduced in Ref.~\cite{cheng2025weaklysupervisedanomalydetection}, which proceeds in three ML-driven stages.
The first stage embeds the collider observables of each event into a latent space.
The second stage uses generative models to estimate the background distribution in this latent space, interpolating from the Higgs boson mass sidebands into the signal region.
The third stage trains classifiers to distinguish the observed data from the estimated background; those classifiers are sensitive to a potential signal and their output is used to select a region enriched in non-background-like events.
Figure~\ref{fig:analysis_strategy} provides a diagram of the three-stage analysis strategy.
This work introduces entirely new embedding strategies for the first stage, which in turn shape the background estimation and classification performed in the defined latent space.
In addition, it introduces a new inference framework from which both signal-agnostic and signal-specific limits are derived, broadening the sensitivity claims towards those expected from a full, dedicated search.
\begin{figure}[tbh]
\centering
\includegraphics[width=1\textwidth]{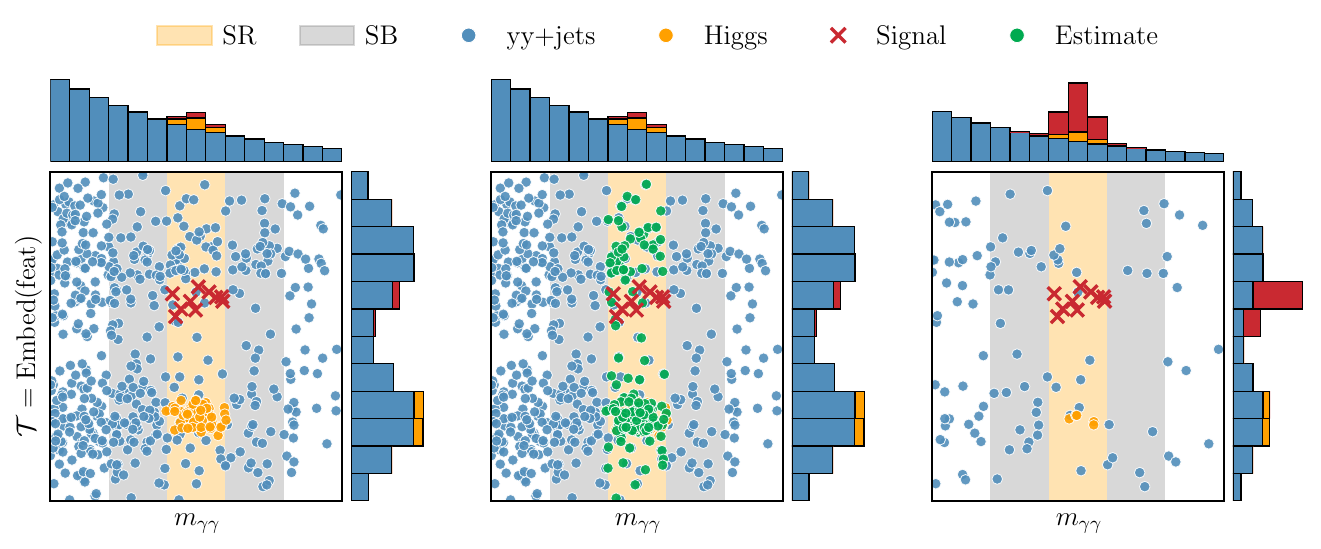}
\caption{\label{fig:analysis_strategy} Diagram of the HAXAD analysis strategy, proceeding through its three ML-driven stages, shown left to right: feature embedding, CATHODE-based background estimation, and CWoLa weak classification. In each panel, the horizontal axis shows \myy\ and the vertical axis shows the embedded feature set $\tau$, in which the signal region (SR, orange band) and Higgs boson mass sidebands (SB, grey bands) are defined. Throughout, $\gamma\gamma$+jets background events are shown in blue, SM Higgs boson events in orange, and potential signal events as red crosses. The left panel illustrates the definition of the SR and SB in the embedded feature space, populated by background, Higgs boson, and potential signal events. The center panel illustrates the background estimation: the estimated background (green), obtained by interpolating the non-resonant component from the SB into the SR and adding the resonant SM Higgs boson contribution directly, is compared to the true background and Higgs boson populations in the SR. The right panel illustrates the effect of the weakly supervised classifier selection, which suppresses the background and Higgs boson populations relative to the signal, enriching the selected region in potential signal events.}
\end{figure}


\subsection{Embedding Methods}
\label{sec:analysis:embed}

In previous work, a \gls{VAE}~\cite{Kingma:2013hel} trained on the sideband data was used to embed detector observables into a new basis better suited to the CATHODE-based background generation~\cite{Hallin:2021wme} while preserving signal sensitivity. 
This encoder leveraged nine features: the \pt\,of the diphoton system, the angular separation between the two photons, $p_{\mathrm{T}}^{j_1}$ and $p_{\mathrm{T}}^{j_2}$ of the leading and subleading jets, the invariant mass $m_{jj}$ and angular separation of the leading dijet system, a binary flag indicating the presence of at least one electron or muon, the scalar sum of the transverse momenta of jets with $p_{\mathrm{T}} > 30\,\mathrm{GeV}$, and the missing transverse momentum $E_{\mathrm{T}}^{\mathrm{miss}}$.
Although this encoder design has shown sensitivity to both signal models tested in Ref.~\cite{cheng2025weaklysupervisedanomalydetection}, this work considers a wider range of signal models to improve upon the previous approach.
Since the \gls{VAE} model is trained on the \gls{SB} data only, which is dominated by diphoton events with very few $b$-jets and leptons present, the encoder tends to ignore those features, which makes it insensitive to processes like $\tilde\chi_1^+(H\ell^+) \tilde\chi_1^-(H\ell^-)$ and \ttFCNCall.
In this study, two new embedding methods are considered which both improve the performance of the entire HAXAD pipeline.

\subsubsection{Unsupervised}

The first embedding method is a fully unsupervised \gls{VAE}, trained on MC simulations of the continuum diphoton background together with the \gls{SM} Higgs boson processes.
The training sample consists of the three continuum components ($\gamma\gamma$+jets, $t\bar{t}\gamma\gamma$, and $\ell\ell\gamma\gamma$+jets) with diphoton invariant mass in the full analysis window $105 < m_{\gamma\gamma} < 145\,\mathrm{GeV}$, together with the four \gls{SM} Higgs production modes (ggH, VBF, $VH$, and $t\bar{t}H$) restricted to the signal region $120 < m_{\gamma\gamma} < 130\,\mathrm{GeV}$. In total, 7.9M simulated training samples are used to train the model, with 5.0M events coming from the dominant $\gamma\gamma$ background and 2.9M events from the \gls{SM} Higgs boson processes.
The model acts on 19 event-level features, which include the same nine input features used in the previous \gls{VAE} model plus \pt, $\eta$, $\phi$, and flavor of the leading two leptons, and a $b$-tag identification flag for the leading two jets.
Positive-definite kinematic quantities are log-transformed and all continuous features are standardized to zero mean and unit variance, while binary flags are passed through unchanged.
The encoder and decoder are symmetric \glspl{MLP} with two hidden layers of 128 nodes each and ReLU activations, for a total of about $4.1\times10^{4}$ trainable parameters. The encoder predicts the mean and variance of a diagonal Gaussian posterior over a seven-dimensional latent space, from which the latent vector is sampled via reparameterization. The latent dimensionality is chosen as a compromise between reconstruction fidelity and compression of the input features into a latent space compact enough for the downstream density estimation.
Features associated with absent objects (e.g.\ jets or leptons) are set to zero at the input, corresponding to the mean of the standardized distributions, and a per-event mask restricts the reconstruction loss to the features that are actually measured, with a per-event renormalization such that events with different object multiplicities contribute comparably to the training objective.

The VAE training loss consists of a mean-squared-error reconstruction term for the continuous features and a binary cross-entropy term for the binary flags, combined with the \gls{KL} divergence of the latent posterior with respect to a standard normal prior, weighted by $\beta = 0.1$.
The training sample is randomly split into training and validation sets containing 80\% and 20\% of the events, respectively. The model is trained with a batch size of 4096, using the Adam optimizer~\cite{kingma2017adammethodstochasticoptimization} with an initial learning rate of $10^{-3}$ and cosine annealing to $10^{-6}$ over 300 epochs. No early stopping is applied, and the checkpoint with the lowest validation loss is retained.
For each event, a single latent vector sampled from the encoder posterior serves as the event embedding used by all downstream anomaly detection stages, so that the embedding also carries the width of the learned posterior.

\subsubsection{Semi-supervised}

The second embedding method features the contrastive encoder design introduced in Ref.~\cite{li2026signalawarecontrastivelatentspaces}, which builds a semi-supervised latent space for weakly supervised anomaly detection.
The encoder body is based on a particle transformer architecture~\cite{Vaswani:2017lxt,Qu:2022mxj} and has 1.1M trainable parameters.
The inputs of this encoder are the same as described in Ref.~\cite{li2026signalawarecontrastivelatentspaces}, consisting of both low-level and high-level features. 
The low-level features are 11 objects per event, which contain four small-radius jets, two large-radius jets, two electrons, two muons, and $E_{\mathrm{T}}^{\mathrm{miss}}$.
Each object is described by 12 features, including its four-momentum, $b$-tagging score, two $n$-subjettiness variables $\tau_{32}$ and $\tau_{43}$~\cite{Thaler:2010tr}, and a five-component one-hot encoding indicating whether the object is a small-radius jet, large-radius jet, electron, muon, or $E_{\mathrm{T}}^{\mathrm{miss}}$.
Padding with zeros is applied for missing objects (no muons in an event, for example) and features ($n$-subjettiness for leptons, for example).
The high-level features contain $\Delta R_{\gamma\gamma}$, $p_{\mathrm{T}\gamma\gamma}$, the number of leptons, the number of jets, the number of $b$-jets, and the number of photons in the event.
The output block of the encoder is inspired by the \gls{VAE}, which predicts a mean and variance from which the latent vector is sampled via reparameterization.
A six-dimensional latent space is used, following the same design as in Ref.~\cite{li2026signalawarecontrastivelatentspaces}.
The smoothing in the \gls{VAE}-inspired output block encourages a Gaussian latent space that is well suited for downstream density estimation and interpolation.
During training, a two-layer \gls{MLP} projection head is applied after the output block and discarded at inference time, a practice that has been shown to improve the quality of the learned embedding~\cite{chen2020simpleframeworkcontrastivelearning}.

Given $N$ events in a batch $I$ with indices $i$ and physics process labels $y_i$, let $z_i$ denote the latent vector from the encoder output block and $Z_i$ the corresponding projection-head output.
Define $A(i) \equiv I \setminus \{i\}$ and $P(i) \equiv \{\, p \in A(i) : y_p = y_i \,\}$ which is the set of all other events in the batch sharing the same label.
The contrastive encoder is trained with the loss:
\begin{equation}
  \label{eq:loss}
  \begin{gathered}
  \mathcal{L}
    = \mathcal{L}_{\mathrm{con}} + \lambda\, \mathrm{KL}\!\left(z \,\middle\|\, \mathcal{N}(0,0.1)\right) \quad \text{with}\\
    \mathcal{L}_{\mathrm{con}}
    = \sum_{i\in I} \frac{-1}{|P(i)|} \sum_{p\in P(i)}
    \log \frac{\exp\!\left(Z_i \cdot Z_p / \tau\right)}
    {\sum_{a\in A(i)} \exp\!\left(Z_i \cdot Z_a / \tau\right)}
  \end{gathered}
\end{equation}
where $\mathcal{L}_{\mathrm{con}}$ is the supervised contrastive learning loss in Ref.~\cite{khosla2020supervised} and the second term is the \gls{KL} divergence with respect to a Gaussian with standard deviation 0.1 for regularization.

During model training, \gls{MC} samples from all processes listed in Table~\ref{tab:samples} are used.
The same label is given to all events in a given process, so models with different mass parameters in the same process share the same label. 
In total, 8M training samples are used to train the model, with 4M events coming from the dominant $\gamma\gamma$ background, and 100k events from each of the \gls{SM} Higgs processes and each of the \gls{BSM} processes at each mass point. 
A total batch size of 16384 is used, with a sampler that ensures uniform representation of each process per batch.
The AdamW optimizer~\cite{loshchilov2019decoupledweightdecayregularization} is employed with an initial learning rate of $2\times10^{-4}$, five epochs of linear warm-up, and cosine annealing~\cite{Loshchilov:2016tgk}.
The training runs for up to 100 epochs. Early stopping is applied when the validation loss does not decrease for 10 epochs.
The contrastive loss temperature is set to $\tau = 0.1$ and the \gls{KL} weight $\lambda$ is set to be 0.1.

By design, the contrastive loss in Eq.~\ref{eq:loss} will pull events belonging to the same process together and push events from different processes apart, thus creating a latent space with clear signal-background separation.
The regularization smoothes the latent space distribution and makes the downstream tasks of background modeling easier.
Although the training process uses signal \gls{MC} and labels for supervision, the contrastive loss itself does not explicitly separate signal from backgrounds but rather distributes events according to their physics properties.
As a result, the encoder remains semi-supervised and is sensitive to potential signal processes not present in the training dataset as shown in Ref.~\cite{li2026signalawarecontrastivelatentspaces}.
Appendix~\ref{app:semisupervised_encoder_extrapolation} explicitly demonstrates the sensitivity of the contrastive encoder to signal models not present in the dataset.
In general, it has been found that adding more signal models in the training data improves the model's ability to perform well on signals that are not present in the training sample.

Compared with the contrastive encoder, the \gls{VAE} model remains fully signal-agnostic, since it is trained without supervision on the non-resonant and \gls{SM} Higgs boson background events only.
The number of parameters in the \gls{VAE} model is also over an order of magnitude smaller than that of the contrastive encoder.
These features offer a basis from which a comparison of supervision level, as well as model size and training time, can be made, aiding analysis-level design choices depending on requirements and constraints. 
In the following text, we will refer to the \gls{VAE} model as the unsupervised encoder and the contrastive encoder as the semi-supervised encoder.  


\subsection{Background Estimation}
\label{sec:analysis:est}

This section describes the estimation of the background in the \gls{SR}, which is used to train the weakly supervised classifier described in \cref{sec:analysis:cwola}.
The estimation is performed in the embedding space provided by the techniques discussed in \cref{sec:analysis:embed}.
The motivation and methodology largely follow Ref.~\cite{cheng2025weaklysupervisedanomalydetection}.
The background comprises two components: a non-resonant component, dominated by continuum QCD diphoton production, and a resonant component from SM Higgs boson production.
The two components are estimated separately, as described in the following.

The contribution of the non-resonant component is assumed to vary smoothly between the \gls{SB} and the \gls{SR}.
Its structure can therefore be learned in the SB and interpolated into the \gls{SR}.
When employed in a real analysis, this approach is entirely data-driven; in this demonstrator study, simulated data are used for simplicity.
The latent features are standardized to zero mean and unit variance using constants derived from the SB.
Then, the distribution of the non-resonant component is modeled with a normalizing flow model~\cite{Papamakarios:2019fms}, which learns the conditional probability density of the latent features given \myy.

The flow model consists of six rational quadratic spline (RQS)~\cite{durkan2019neural} transformation layers with random permutation layers in between.
Each spline uses ten bins on the interval $[-6, 6]$ standard deviations, with linear tails outside.
The spline parameters of each transformation layer are predicted by a \gls{MADE} network~\cite{germain2015made} conditioned on the diphoton invariant mass \myy.
Each \gls{MADE} network is a two-block residual network with eight hidden nodes per latent dimension, corresponding to a hidden layer width of 48 (56) nodes for the semi-supervised (unsupervised) embedding.
The value of \myy{} is projected into the hidden space by a separate linear layer and added to the hidden activations.
The flow model is implemented in PyTorch~\cite{NEURIPS2019_bdbca288}, using the nflows package~\cite{nflows}.

The data from the \gls{SB} are split into a training set with 80\% and a validation set with 20\% of the events.
The flow model is trained by minimizing the negative log-likelihood (NLL) using the Adam optimizer~\cite{kingma2017adammethodstochasticoptimization} with an initial learning rate of 0.008.
An epoch is defined as 1000 iterations with a batch size of 2048.
Whenever the validation loss does not improve for two epochs, the learning rate is halved.
The training terminates once the learning rate has been reduced to $10^{-5}$ of its initial value, or after 500 epochs.
Four independent flow models are trained, each with a different random seed for the weight initialization, the training/validation split, and the data sampling.

The trained flow models are then used to sample events in the \gls{SR}.
First, an exponential function of the form $f(m) = a \exp(-c\,\tilde{m})$, where $\tilde{m}$ is the diphoton invariant mass rescaled to a normalized coordinate, is fitted to the \myy{} spectrum.
In this demonstrator study, the fit is performed on simulated data of the non-resonant background component over the entire \myy{} spectrum; in an analysis of experimental data, it would be performed on the recorded spectrum in the SB only.
The fit is binned, using 500\,MeV bins and a weighted least-squares minimization.
Diphoton invariant masses are then drawn from the fitted spectrum by rejection sampling, restricted to the SR, and used as conditional values for the flow models.
Using these sampled \myy{} values, each of the four flow models generates 500{,}000 events, giving a total of 2{,}000{,}000 events.
The events are oversampled to reduce the statistical uncertainties of the sampled template.
The generated latent features are transformed back from the standardized to the original scale.
All generated events receive the same weight, chosen such that the total weight equals the integral of the fitted exponential over the SR mass range.

The resonant component consists of SM Higgs boson production with the Higgs boson decaying to two photons, and is modeled directly with simulations.
The ggH, VBF, VH, and ttH production modes are included and scaled to their predicted cross sections and the assumed integrated luminosity; more information about the sample generation is given in \cref{subsec:samples}.
In this demonstrator, the simulated events are statistically independent of the pseudo-data but follow exactly the same distributions.
This simplification does not hold exactly when applied to recorded data; however, we expect Higgs boson simulations at collider experiments to be accurate enough for this approximation to be justified.

Together, the non-resonant and resonant components form the background estimate used to train the weakly supervised classifier.
The closure of the background estimation is presented in Appendix~\ref{app:inputs} (\cref{fig:unsup_feature_distribution,fig:semisup_feature_distribution}), where the latent-feature distributions of the estimated background are compared to those of the pseudo-data.


\subsection{Weakly Supervised Classification}
\label{sec:analysis:cwola}

After estimating the background in the SR, weakly supervised classifiers are trained to distinguish between the background estimate and the pseudo-data, using the latent features as inputs.
Events from the pseudo-data are assigned the label one and events from the background estimate the label zero.
By the CWoLa theorem~\cite{Metodiev:2017vrx}, this classifier converges to the optimal classifier between signal and background, given that the background estimate accurately models the background in the data and in the limit of an infinite amount of training data.
As for the previous stages, the methodology largely follows Ref.~\cite{cheng2025weaklysupervisedanomalydetection}.

Following other AD strategies~\cite{Finke:2023ltw,Freytsis:2023cjr}, we employ Boosted Decision Trees (BDTs)~\cite{Friedman:2001wbq}, as they have been observed to perform well in a regime of sparse data and can handle potentially uninformative features.
The number of estimators is set to 50 and the maximum tree depth is limited to five.
The BDTs are trained with a weighted binary cross-entropy loss and a learning rate of 0.01.
The data are split with a 5-fold cross-validation to prevent overfitting: each BDT is trained on three fifths of the data, validated on one fifth, and applied to the remaining fifth.
Early stopping with a patience of five boosting rounds is used, as evaluated on the validation loss.
The classifiers are implemented using the XGBoost package~\cite{DBLP:journals/corr/ChenG16}.

To reduce statistical uncertainties, 20 such cross-validated classifiers are trained: five against each of four independent realizations of the embedding and background estimate, and all differing in their random seed, which determines both the cross-validation split and the BDT initialization.
The classifiers are grouped into five ensembles of four at random, irrespective of the underlying realization of the embedding and background estimate, and the scores are averaged within each ensemble.
For each ensemble, a selection retaining the highest-score events is applied to enrich a potential signal.
The resulting five \myy{} spectra are averaged to obtain the nominal distribution used in the statistical inference described in \cref{sec:inference}, which also accounts for their variation.
The performance of the classifiers is shown in \cref{subsec:sensitivity} in the context of the full HAXAD analysis chain.

\subsection{Inference}
\label{sec:inference}

The statistical interpretation of the selected data is based on a binned maximum-likelihood fit to the \myy{} spectrum of the events passing the classifier selection of \cref{sec:analysis:cwola}.
A model-independent upper limit on the number of signal events is derived first.
This limit constrains any anomalous production of Higgs bosons in association with other objects without reference to a specific signal hypothesis, and constitutes the primary result of the anomaly detection search.
It is subsequently translated into cross section limits for specific benchmark signal models.

For each classifier ensemble, the selection retains the 0.05\% of pseudo-data events in the \gls{SR} with the highest classifier score.
The resulting score threshold is applied unchanged to the \gls{SB} data, to the \gls{SM} Higgs simulation from which the resonant contribution to the fit model is determined, and to the continuum simulation used in the spurious-signal test described below.
The maximum-likelihood fit is performed on the combined \myy{} spectrum of \gls{SB} and \gls{SR} events, covering 105--145\,GeV in 20 bins of 2\,GeV.
About 264 events are retained in the \gls{SR}.

The five post-cut \myy{} spectra of the classifier ensembles are averaged to form the nominal spectrum, as described in \cref{sec:analysis:cwola}.
Since the ensembles select overlapping subsets of the same events, the averaging suppresses the stochastic variation of the classifier training rather than the statistical fluctuations of the data.
The statistical power of the averaged spectrum therefore corresponds to that of a single ensemble.

The parameter of interest is the number of signal events $N$ of an additional Higgs-boson-like resonance at 125\,GeV.
Because the anomaly is assumed to appear in the objects produced in association with the Higgs boson rather than in the diphoton system itself, the resonance mass is fixed and no scan over \myy{} is performed.
This choice avoids the trials penalty (look-elsewhere effect)~\cite{Gross:2010qma} that would be introduced by testing multiple mass hypotheses.
The upper limit on $N$ is model independent in the sense that it constrains any signal producing a \gls{SM}-like $H\to\gamma\gamma$ decay in the selected region.
Its translation into cross section limits for specific signal models must account for the dependence of the selection efficiency on the signal itself and is described at the end of this section.

Both the signal and the resonant \gls{SM} Higgs background are modeled by a double-sided Crystal Ball function, a Gaussian core with power-law tails on both sides, as commonly used in $H\to\gamma\gamma$ analyses~\cite{ATLAS:2022tnm}.
Its parameters are determined from the \gls{SM} Higgs simulation passing the score cut.
The five per-ensemble \myy{} spectra are resampled with replacement 1000 times following the bootstrapping approach, each replica is averaged and fitted, and the mean of each fitted parameter over the replicas is taken as its nominal value.

The reconstructed \myy{} shape is determined by the $H\to\gamma\gamma$ decay kinematics and the detector resolution, and is therefore insensitive to the production mechanism of the Higgs boson.
The same shape consequently describes both the \gls{SM} Higgs background and the \gls{BSM} signals targeted by this analysis.
The \gls{SM} Higgs yield $N_H$ is fixed to its post-cut expectation obtained from the same bootstrap procedure.
Because $N_H$ is held fixed, the parameter $N$ measures any excess of Higgs-boson-like events above the \gls{SM} Higgs expectation.
In particular, if the classifier selects more \gls{SM} Higgs events than predicted by the simulation, the difference appears in the fitted value of $N$.
The residual dependence of the limits on the fixed shape and normalization values is expected to be small compared to the background-modeling uncertainty described below, and these parameters are therefore held constant in the fit.

The non-resonant continuum background is described by an analytic function whose normalization and shape are determined directly in the final fit, following the functional-form approach recommended for smoothly falling diphoton mass distributions~\cite{ATL-PHYS-PUB-2020-028}.
An inadequate choice of function can bias the fitted signal yield.
This bias is quantified with a spurious-signal test~\cite{ATLAS:2022tnm}, in which a signal-plus-background fit is performed on a background-only sample.
The signal yield extracted by this fit, denoted $N_{\mathrm{sp}}$, would vanish for an unbiased background model, so its magnitude measures the bias of the candidate function.

The background-only sample is the simulated continuum diphoton sample after the per-ensemble score thresholds, averaged over the five classifier ensembles.
Performing the test on a sample selected identically to the data ensures that any sculpting of the continuum shape by the classifier selection is reflected in the test.
The candidate functions consist of power laws, exponentials of polynomials, and Bernstein polynomials.
For each candidate, the fitted yield is evaluated for hypothesized resonance masses between 121 and 129\,GeV.
This scan probes the fit bias anywhere the signal shape has support, while the limit fit itself is always performed at 125\,GeV.

A candidate function is accepted if it satisfies two criteria.
First, the largest absolute yield must satisfy $\max|N_{\mathrm{sp}}| \leq 0.2\,\delta_{125} + 2\,\sigma_{125}$, where $\delta_{125}$ is the expected statistical uncertainty of the fitted yield at a resonance mass of 125\,GeV and $\sigma_{125}$ is the statistical uncertainty arising from the limited size of the simulated sample.
Second, the fit quality must be acceptable, with $\chi^2/\mathrm{ndf} < 3$, a loose requirement that rejects only pathological fits.
Among the accepted candidates, the function with the fewest free parameters is selected.
Since a single set of pseudo-data underlies the model-independent limit, the test is performed once and results in a single choice of background function for this study.
The chosen function is an exponential, which coincides with the form used for the mass sampling in \cref{sec:analysis:est}.

The associated uncertainty $\delta_{\mathrm{spur}}$ is the 95th percentile of the $\max|N_{\mathrm{sp}}|$ distribution over 1000 bootstrap replicas formed by resampling the five per-ensemble continuum spectra, with the choice of function kept fixed.
The bootstrap thereby propagates the variation of the spurious signal across the classifier ensembles.
Taking the 95th percentile instead of the mean is a conservative choice.
In an analysis of recorded data, the same test would be performed on a background-only template with much larger statistics than the data, as done in ATLAS $H\to\gamma\gamma$ measurements~\cite{ATLAS:2022tnm}.

The likelihood of the binned \myy{} spectrum is
\begin{equation}
\label{eq:likelihood}
  \mathcal{L}(N, \boldsymbol{\theta}) =
  \prod_{i=1}^{n_\mathrm{bins}} \mathrm{Pois}\!\left( n_i \,\middle|\,
  \left( N + N_H + \theta_{\mathrm{spur}}\,\delta_{\mathrm{spur}} \right) f^{\mathrm{sig}}_i
  + B\, f^{\mathrm{bkg}}_i(c) \right)
  \cdot \mathcal{G}(\theta_{\mathrm{spur}}),
\end{equation}
where $n_i$ is the content of bin $i$ of the averaged spectrum, $f^{\mathrm{sig}}_i$ and $f^{\mathrm{bkg}}_i$ are the fractions of the signal and continuum shapes in bin $i$, $N_H$ is the fixed yield of the resonant \gls{SM} Higgs background, which shares the signal shape with its own normalization, and the nuisance parameters $\boldsymbol{\theta}$ consist of the freely floating continuum yield $B$ and slope $c$ together with the spurious-signal parameter $\theta_{\mathrm{spur}}$.
The spurious-signal term enters the signal yield additively.
The parameter $\theta_{\mathrm{spur}}$ is the only constrained nuisance parameter, with a unit-Gaussian constraint term $\mathcal{G}$.
No theory uncertainties are assigned in this demonstrator study.  

In summary, the statistical uncertainty of the data enters through the Poisson terms.
The background-modeling bias and its variation across the classifier ensembles enter through $\delta_{\mathrm{spur}}$.
The stochasticity of the pseudo-data sampling and of the classifier training enters the model-dependent interpretation, in which the classifiers are retrained for every injected signal.
It is quantified by repeating the signal-injection procedure described below over independent pseudo-data samplings.
No further uncertainties are included.

Upper limits on $N$ at the 95\% confidence level (CL) are derived with the $\mathrm{CL_s}$ method~\cite{Read:2002hq}, which protects against excluding signals to which the analysis has little sensitivity.
The $p$-values are computed from the one-sided profile-likelihood-ratio test statistic in the asymptotic approximation~\cite{Cowan:2010js}.
The median expected limit and its $\pm1\sigma$ and $\pm2\sigma$ bands are obtained from the background-only Asimov dataset.
The expected limits quantify the sensitivity of the analysis.
The observed limit is obtained from the fit to the pseudo-data and reflects the statistical fluctuations of the particular pseudo-data set analyzed.
The identical statistical treatment, including a dedicated spurious-signal test, is applied to each selection of the cut-based comparison analysis.

The upper limit on $N$ holds for any signal with a Higgs boson in its final state, but its translation into a cross section limit does not, because the selection efficiency of a weakly supervised classifier depends on the amount of signal present in the training data, as first addressed in the CMS model-agnostic dijet search~\cite{cmsad_2025}.
If little signal is present, the training converges to an essentially random classifier and few signal events pass the selection.
If the signal is abundant, the classifier identifies its phase space and the selection efficiency rises accordingly.
The limit on $N$ therefore cannot be translated into a cross section by dividing by a single, signal-independent efficiency.

Instead, following the efficiency-calibration procedure introduced in Ref.~\cite{cmsad_2025}, the expected signal yield after selection is calibrated as a function of the signal cross section.
For each signal model, simulated signal events are injected into the pseudo-data at cross sections between 0 and 8\,fb.
The weakly supervised classifiers are retrained at each injection point, while the embedding model and the background estimate are kept fixed.
The fraction of pseudo-data events retained by the score cut is held fixed during the scan, so that the threshold adapts to the injected signal exactly as it would to a signal present in recorded data.

The number of injected signal events passing the selection is determined from the generator-level labels of the pseudo-data and averaged over the five classifier ensembles.
In an analysis of recorded data, this calibration would instead rely on simulated signal samples.
The scan is repeated with three independent samplings of the pseudo-data.
Their mean defines the nominal injection curve.
Their spread defines an envelope reflecting the statistical variation of the pseudo-data sampling together with the classifier retraining.
This envelope does not constitute a frequentist confidence band.

The model-dependent cross section limit is defined as the cross section at which the injection curve crosses the model-independent limit on $N$, which is obtained from the pseudo-data before any signal injection.
The crossing point is found by piecewise-linear interpolation.
The uncertainty of the limit is evaluated from the crossings of the injection curve and its envelope with the expected-limit bands.
At the crossing, the excluded and the injected signal yields coincide by construction.
The quoted value is therefore the smallest cross section that the analysis would have excluded had the signal been present in the data.

This calibration is valid in the absence of a significant signal in the analyzed data; Ref.~\cite{cmsad_2025} also showed that it retains proper coverage for underlying excesses below approximately three standard deviations.
\Cref{fig:inj_scan} illustrates the procedure with a demonstrative pseudo-data scan.
The injection curve grows with the injected cross section, while the model-independent limit on $N$ and its expected bands do not depend on it.
Their crossing, indicated by the vertical line, defines the model-dependent cross section limit.
The scan shown is purely illustrative and does not correspond to any specific signal model.
The model-dependent limits obtained for the benchmark signal models are presented in \cref{subsec:limits}.

\begin{figure}[tbh!]
  \centering
  \includegraphics[width=0.7\textwidth]{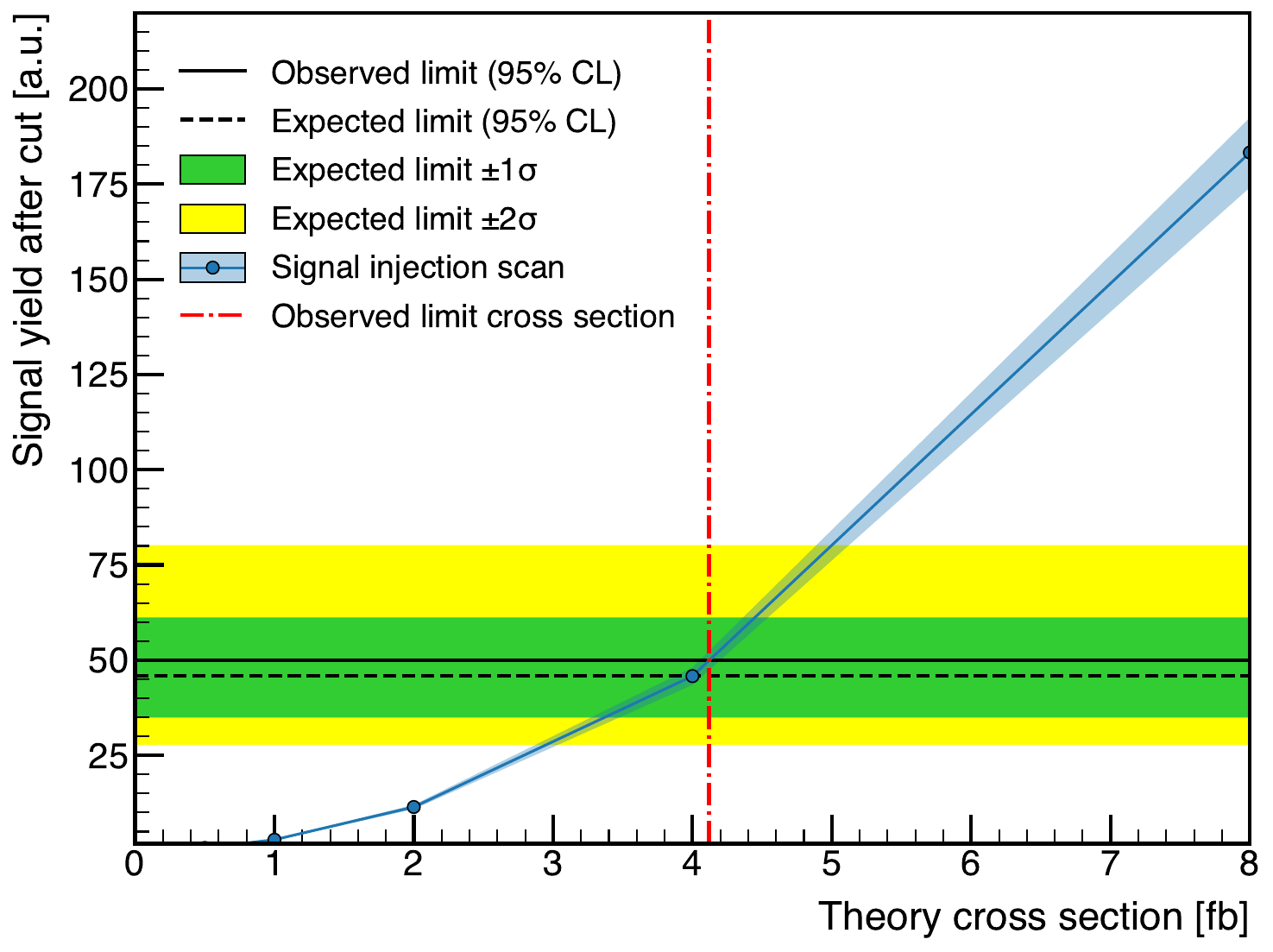}
  \caption{Illustration of the signal-injection limit-setting procedure with a pseudo-data scan that does not correspond to any specific signal model. The number of injected signal events passing the selection, in arbitrary units, is shown as a function of the injected cross section, together with the model-independent upper limit on $N$ and its expected bands. The model-dependent cross section limit is given by the crossing of the injection curve with the observed limit (dash-dotted line).}
  \label{fig:inj_scan}
\end{figure}

\section{Results}
\label{sec:results}

This section presents the performance of the HAXAD analysis in three categories.
First, the sensitivity of the weakly supervised classifiers is quantified in \cref{subsec:sensitivity}.
Second, model independent limits are shown with and without signals injected in data in \cref{subsec:limits_model_indep}.
Third, model dependent limits on the signal cross sections are derived in \cref{subsec:limits}.
The results from the two embedding models are compared with each other and to a collection of cut-based selections; the classifier sensitivity is in addition compared to the initial version of the HAXAD strategy (v1)~\cite{cheng2025weaklysupervisedanomalydetection}.
The latter is inspired by Ref.~\cite{ATLAS:2023omk} and includes requirements on the jet multiplicity ($\geq 6j$, $\geq 8j$), the $b$-jet multiplicity ($\geq 2b$, $\geq 3b$), the lepton multiplicity ($\geq 1l$, $= 2l$), $E_{\mathrm{T}}^{\mathrm{miss}}$ ($> 100$, $> 200$, $> 300\,\mathrm{GeV}$), and $H_{\mathrm{T}}$ ($> 1000$, $> 1500\,\mathrm{GeV}$), as well as cuts on the presence of a lepton and a $b$-jet ($lb$), and the presence of a hadronically decaying top-quark candidate ($top_{had}$).

\subsection{Classifier Sensitivity}
\label{subsec:sensitivity}

This section presents the sensitivity of the trained classifiers using the exemplary benchmark models introduced in \cref{subsec:samples}.
Sensitivity to a given signal model is quantified by the \gls{SIC}, defined as the ratio of the significance after the classifier selection to the inclusive significance before the classifier selection, $\mathrm{SIC} = (S_\mathrm{sel}/\sqrt{B_\mathrm{sel}})/(S/\sqrt{B})$, where $S$ ($B$) denotes the number of signal (background) events in the \gls{SR}.

Figure~\ref{fig:sic_1sigma} compares the \gls{SIC} of four approaches: the classifiers based on the semi-supervised, unsupervised, and HAXAD v1 encoders, and the best cut of the cut-based evaluations.
The comparison is performed at classifier working points that keep $\varepsilon_B = 0.05\%$ of the estimated background in the SR.
Signals are injected such that the inclusive significance before the classifier cut is $S/\sqrt{B} = 1$, with the background yield B in the \gls{SR} of 527,580.
This initial significance corresponds to an $S/B$ ratio of 0.14\% in the \gls{SR}.
The measurement is repeated on three statistically independent pseudo-data sets, introduced in \cref{subsec:samples}, each evaluated with the five classifier ensembles described in \cref{sec:analysis:cwola}.
Shown are the mean and spread across the three pseudo-data sets, including the spread of the five classifier ensembles.
The uncertainty thereby accounts for statistical fluctuations of the data as well as uncertainties from the initialization and training of the ML models.

\begin{figure}[tbh!]
  \centering
  \includegraphics[width=\textwidth]{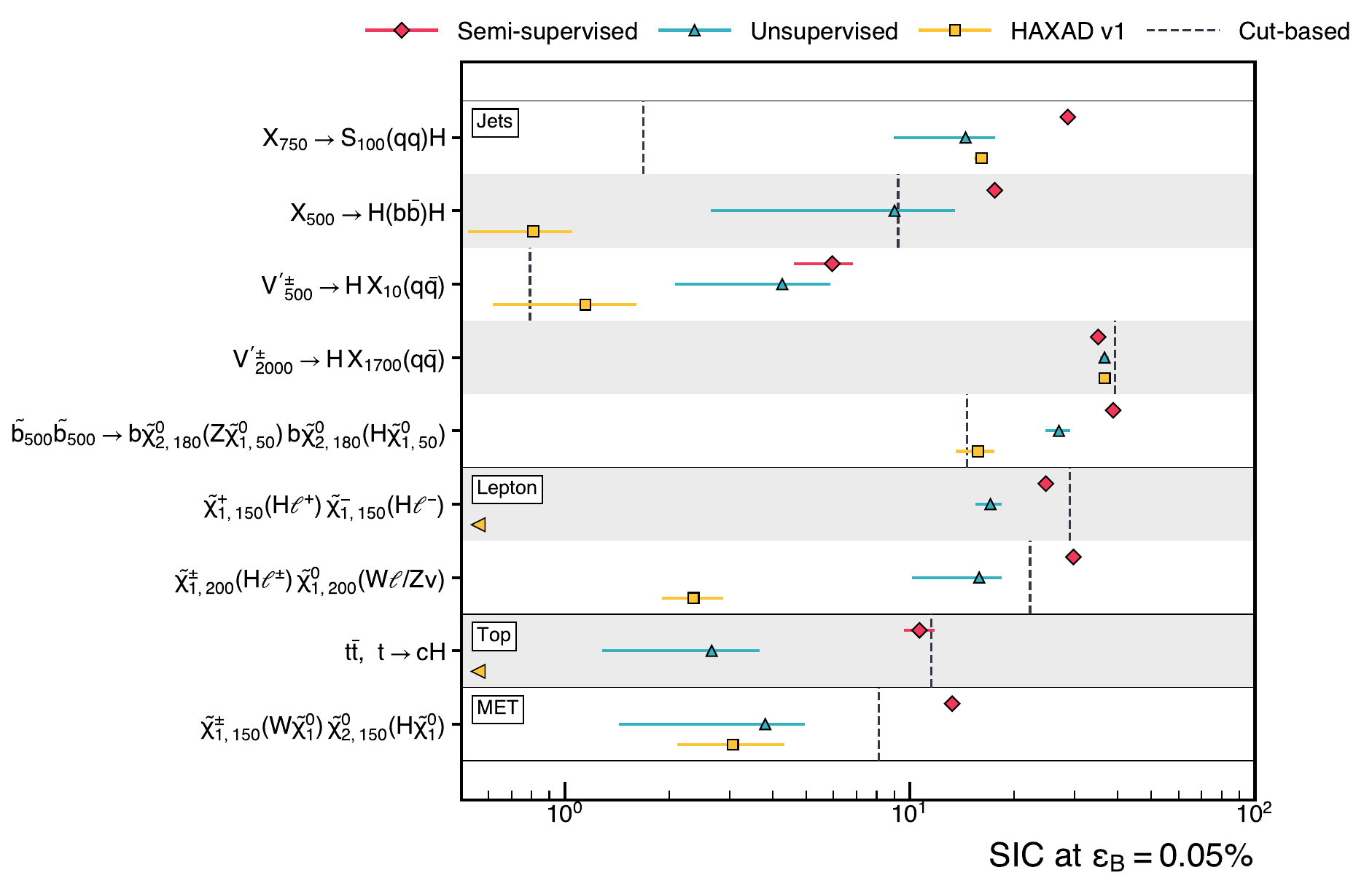}
  \caption{\label{fig:sic_1sigma} \gls{SIC} at $\varepsilon_B = 0.05\%$ with initial signal significance $S/\sqrt{B} = 1$. Four approaches are included: the HAXAD method considering the semi-supervised, unsupervised, and v1 embedding models, and the cut-based analysis. Left-pointing arrows indicate entries below the displayed range.}
\end{figure}

Both the semi-supervised and the unsupervised embeddings are sensitive to a broad range of signal models, whereas a comparable coverage with selection cuts requires the combination of many dedicated selections.
The semi-supervised and unsupervised embeddings are both always on par with or better than the previous HAXAD v1 encoder, with the semi-supervised almost always delivering the best performance.
They largely outperform HAXAD v1 on models whose signatures are rare in the pseudo-data, most notably leptons or top quarks in the final state.
A likely explanation is that such rare signatures are mostly ignored by the purely data-trained v1 embedding, while training on simulated background with sufficient statistics for these processes preserves them.
Finally, for several signal models, HAXAD outperforms even the best individual cut-based region.
Only for models with a clear one-dimensional signature, such as the $V'$ or the leptonic signals, do individual cuts achieve a performance close to the machine-learning-driven strategies.

Even if the semi-supervised method explicitly uses signal models in training, its sensitivity extrapolates to signals not included in the training, as demonstrated in Ref.~\cite{li2026signalawarecontrastivelatentspaces} and in this analysis context in Appendix~\ref{app:semisupervised_encoder_extrapolation}. 
Nevertheless, its sensitivity remains larger for the signal models which are used in the training.
In contrast, the unsupervised encoder is entirely signal agnostic.
Its weaker and less stable performance can be traced back to its training objective, introduced in \cref{sec:analysis:embed}.
Since the \gls{VAE} is trained purely to reconstruct its inputs, there is no guarantee that the features separating a given signal from the background survive the compression into the latent space~\cite{Finke:2021sdf,Fraser:2021lxm}.
In particular, sparse features such as leptons or $E_{\mathrm{T}}^{\mathrm{miss}}$ contribute little to the reconstruction loss and are prone to being compressed away, which explains the performance gaps for signals with such signatures.
Moreover, the reconstruction objective only weakly constrains the shape of the latent space, so independent trainings can converge to noticeably different representations, which is reflected in the larger uncertainty bands compared to the semi-supervised encoder.

\subsection{Model Independent Limits}
\label{subsec:limits_model_indep}

Upper limits on the BSM signals are evaluated to benchmark the sensitivity of the analysis, with the conventional cut-based analysis included for comparison.
Limits are projected to a luminosity of 470\,\ifb, which roughly corresponds to the combined datasets taken by ATLAS during Runs~2 and Run~3 of the LHC.

First, model-independent limits on the number of signal events $N$ are set at the 95\%~CL via the statistical inference procedure described in \cref{sec:inference}.
These limits only assume a generic signal that produces a Standard-Model-like Higgs boson decaying to two photons.
They are quoted as a fiducial cross section $N/L$, with the integrated luminosity $L$, and no signal efficiency is applied.
The projection onto the cross section of a specific signal model follows in \cref{subsec:limits}.

The cut-based comparison evaluates 13 selections, so the probability of observing a deviation in at least one of them exceeds that of a single measurement.
Its significances are therefore corrected for this multiplicity by multiplying the local $p$-value by a trials factor before converting it into a significance~\cite{Gross:2010qma}.
The 13 selections are not independent, since several are built from the same distribution, such as $\geq 2b$ and $\geq 3b$.
A trials factor of six is applied, corresponding to the six distributions underlying the selections: the $b$-jet multiplicity, the jet multiplicity, $H_{\mathrm{T}}$, $E_{\mathrm{T}}^{\mathrm{miss}}$, the lepton multiplicity, and the $lb$ and $top_{had}$ requirements.
This choice is intermediate between the two extremes: treating the six distributions as fully independent maximizes the penalty attributed to them, while counting six groups rather than 13 selections minimizes the penalty attributed to the several selections applied to the same distributions.
For the HAXAD results, no trials factor is applied, since an analysis of recorded data would use a single classifier configuration and report a single limit.
The two encoder configurations are shown here only to compare the embedding strategies on pseudo-data.
This is one of the major advantages of a weakly supervised multivariate method like HAXAD: the classifier compresses many features into a single test, probing a deviation in many directions of phase space without performing many experiments.

\Cref{fig:discovery_model_indep} shows the expected and observed model-independent limits for HAXAD with the semi-supervised and unsupervised embedding strategies, compared to the cut-based selections.
\Cref{fig:discovery_model_indep:without} shows the limits on background-only pseudo-data.
All approaches reach limits of comparable size, around XX\,\ifb, on the dataset without any signal.
\Cref{fig:discovery_model_indep:with} shows the expected and observed limits on pseudo-data with an injected signal using \WNHyyN{150} at an initial significance of 0.5$\sigma$.
Once this exemplary signal is present, HAXAD reaches a significance above 5$\sigma$, exceeding the XX$\sigma$ of the most sensitive cut-based selection.

\begin{figure}
    \centering
    \begin{subfigure}{0.85\textwidth}
        \centering
        \includegraphics[width=\textwidth, page=1]{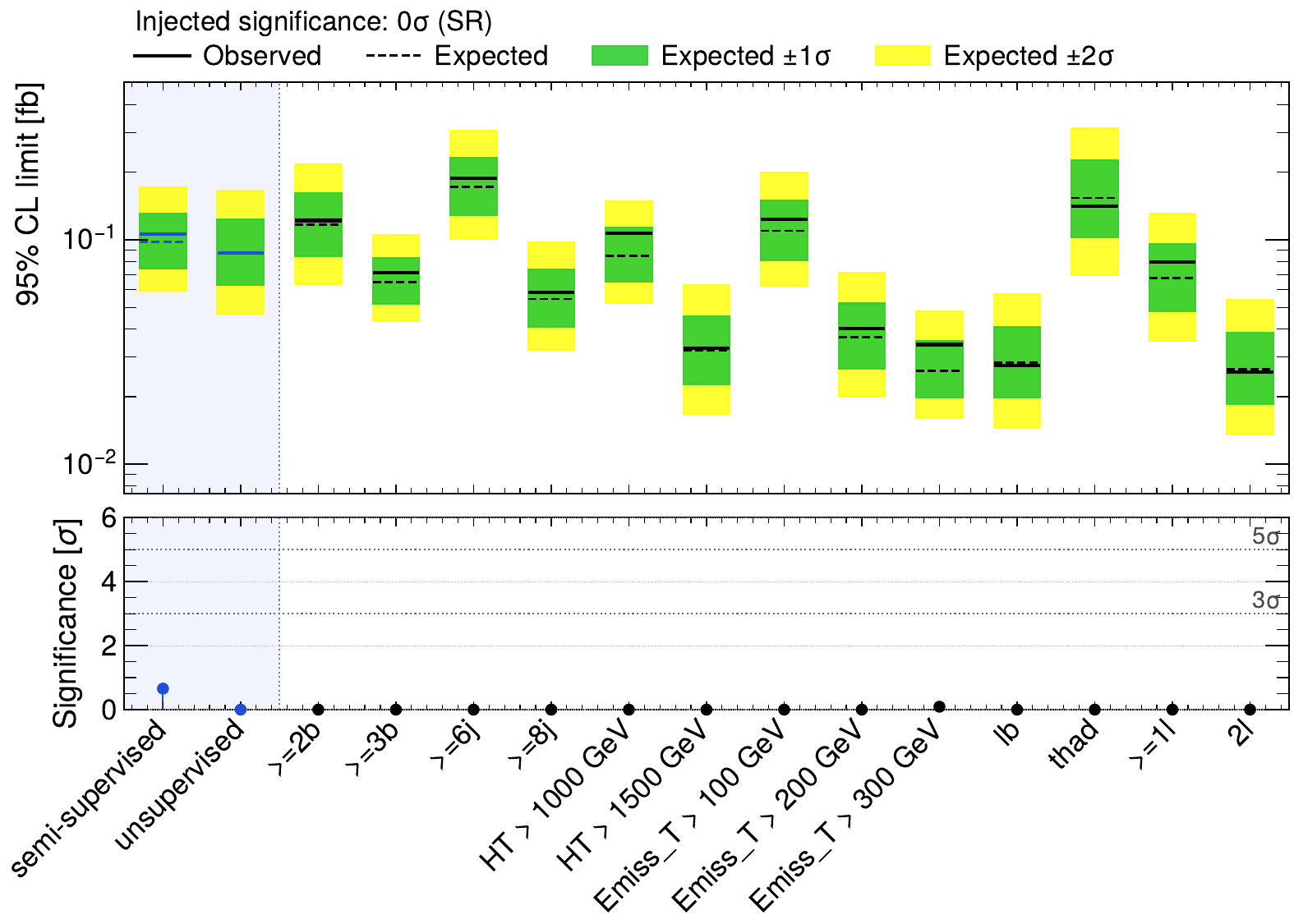}
        \caption{\label{fig:discovery_model_indep:without}Without injected signal}
    \end{subfigure}
    \par\bigskip
    \begin{subfigure}{0.85\textwidth}
        \centering
        \includegraphics[width=\textwidth, page=1]{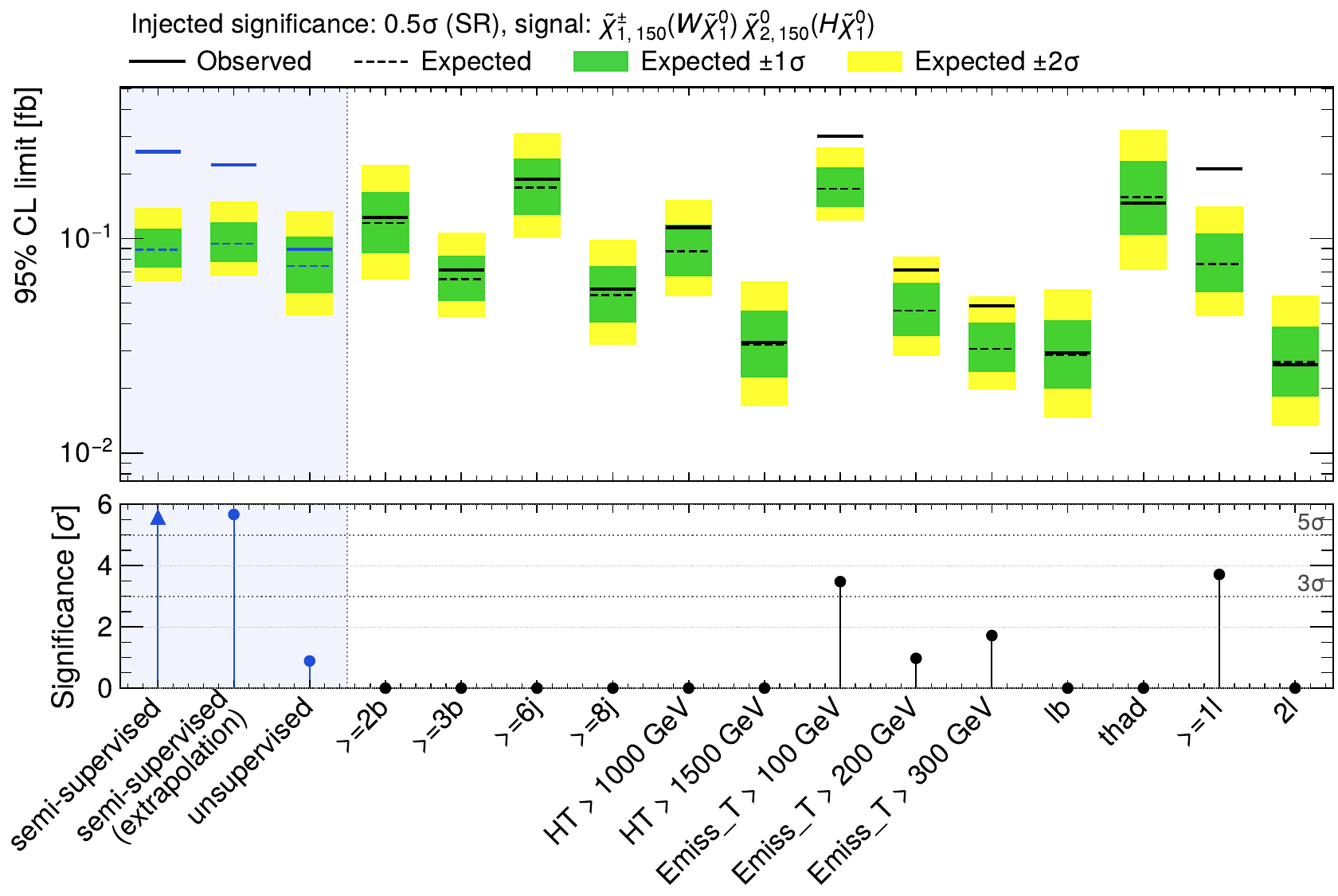}
        \caption{\label{fig:discovery_model_indep:with}With injected signal}
    \end{subfigure}%
    \caption{\label{fig:discovery_model_indep} Cross section limit at 95\%~CL and statistical significance (a) without injected signal and (b) with injected signal using \WNHyyN{150} with an initial significance of 0.5$\sigma$. The unsupervised and semi-supervised embedding approaches are compared to several cut-based regions defined in the ATLAS Run~2 search~\cite{ATLAS:2023omk}. For (b), the semi-supervised limits are shown for the cases of having all (plain) and no (extrapolation) \WNHyyNall~signal models in the training. Limits are projected to a total integrated luminosity of 470\,\ifb.}
\end{figure}


\subsection{Model Dependent Limits}
\label{subsec:limits}

The model-independent limit on $N$ is projected onto the cross section of a specific signal through the signal efficiency $\epsilon$, defined as the number of signal events entering the signal region divided by the number in the full phase space considered by the analysis.
The latter corresponds to the fiducial region defined by the common pre-selection, the trigger emulation, and the diphoton mass window described in Section~\ref{subsec:samples}. The limits are quoted for the specific $H \to \gamma\gamma$ decay channel, with its branching ratio included in the cross section.

For the conventional cut-based analysis the signal efficiency is a constant, measured directly on the signal MC samples, and the cross section limit is obtained as $N/(L\,\epsilon)$ with $L$ being the integrated luminosity.
No signal theory uncertainty is assigned, and the experimental uncertainties are already fully propagated in the preceding model-independent limit step.

For the HAXAD analysis the signal efficiency is not constant: it depends on the amount of BSM signal present in the data, because the weak classifiers that define the selection are trained on the data themselves.
The limits are therefore derived with the signal-injection procedure described in \cref{sec:inference}, which determines, for each signal model, the cross section at which the number of injected signal events passing the selection equals the model-independent limit on $N$.

The model-dependent limits are summarized in Fig.~\ref{fig:limits}, where for each signal the most sensitive cut-based region is shown alongside the two HAXAD encoding configurations: the contrastive encoder trained with semi-supervised training and the fully connected autoencoder with fully unsupervised training.
The full set of model-dependent limits for all signal models and mass parameters is provided in Appendix~\ref{app:limits_all} (Fig.~\ref{fig:limits_all}).

\begin{figure}[tbh!]
\centering
\includegraphics[width=\textwidth]{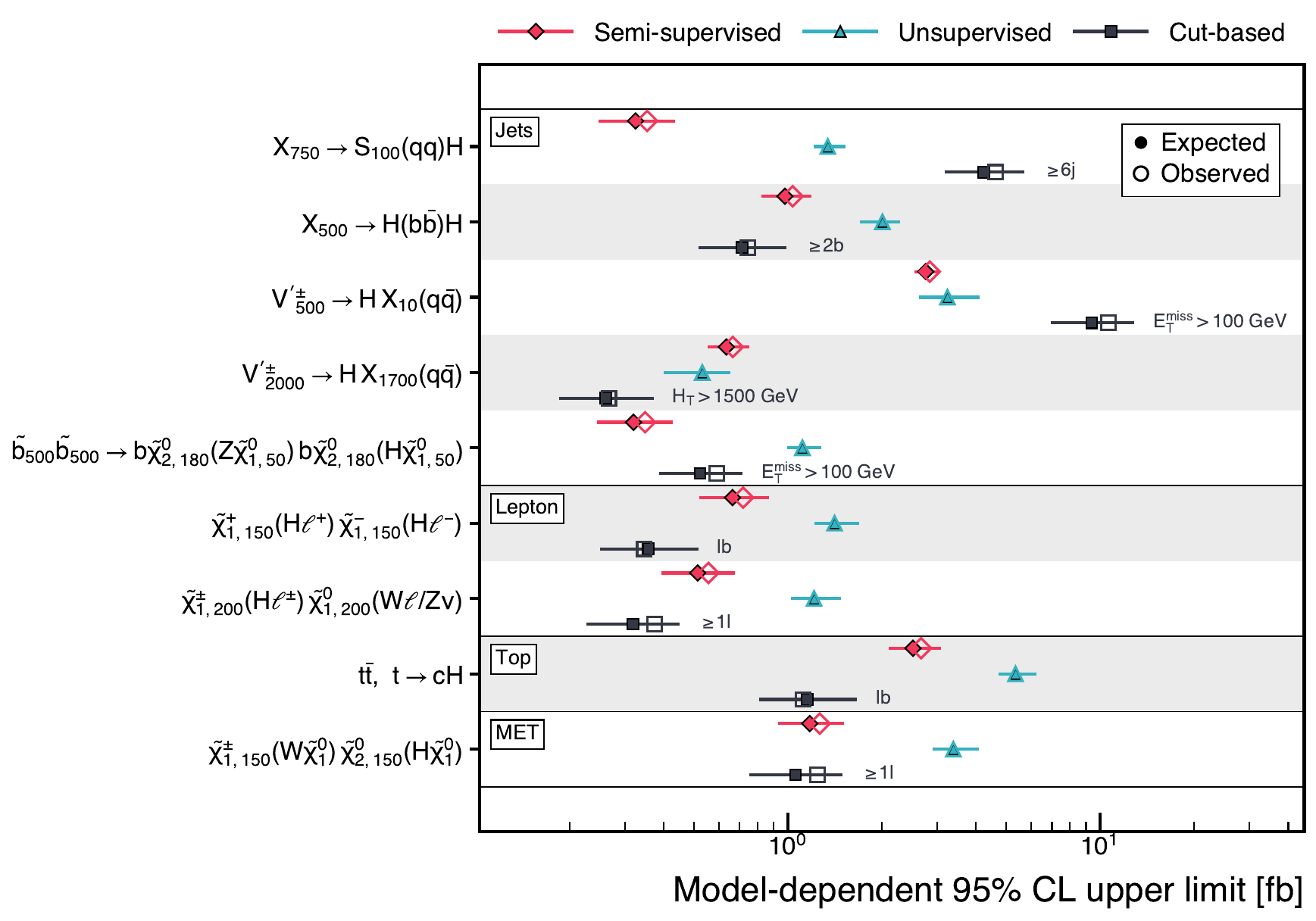}
\caption{\label{fig:limits} Model dependent cross section limit at 95\%~CL for the unsupervised and semi-supervised embedding approaches, compared to the best-performing cut-based region as defined in the ATLAS Run~2 search~\cite{ATLAS:2023omk}. Filled markers with their $\pm1\sigma$ bands show the expected limits and open markers the observed limits. Limits are projected to a total integrated luminosity of 470\,\ifb.}
\end{figure}

\section{Conclusions}
\label{sec:conclusions}

A study is presented extending the performance and utility of the HAXAD approach for new physics searches at colliders. 
The ML-based embedding model is updated, incorporating a comparison of two methods with varying levels of training supervision, both of which provide an improvement in SIC for benchmark signal models compared to existing literature.
The depth and breadth of HAXAD's sensitivity are further quantified with 12 simulated signal models and an interpretation which includes model-independent and model-dependent cross section upper limits.
With respect to a cut-based analysis with regions motivated by a search from the ATLAS Collaboration in a similar phase space, the HAXAD-based analysis strategy achieves stronger limits across a wide variety of signal models, highlighting its anomaly detection capability.
These advances underscore the utility of HAXAD for BSM discovery potential in future collider experiment searches.

\section*{Dataset and code availability}
The simulated dataset used in this work is publicly available on Zenodo~\cite{cheng_2026_23485470}.
The analysis code is publicly available at \url{https://gitlab.cern.ch/haxad/haxad_v2}.

\section*{Declaration of generative AI and AI-assisted technologies}
During the preparation of this work, the authors used Claude Code with the language models \texttt{Opus 4.8} and \texttt{Fable 5} to assist in writing analysis code, structuring the content of the paper, and making language edits. After using these tools, the authors reviewed and edited the content as needed and take full responsibility for the content of the published article.

\acknowledgments
We thank Sarah Demers, Sascha Diefenbacher, and Laura Miller for valuable discussions during the course of this work.
This work is supported by the U.S. Department of Energy under Contract No. DE-AC02-76SF00515 and DE-SC0024518.
This work used resources of the National Energy Research Scientific Computing Center (NERSC), a Department of Energy User Facility using NERSC award HEP-ERCAP 0037461.


\bibliographystyle{JHEP}
\bibliography{biblio.bib}

\FloatBarrier

\appendix
\section{Input and Embedded Features}
\label{app:inputs}
The feature distributions in the latent space constructed by the unsupervised (semi-supervised) encoder are shown in \cref{fig:unsup_feature_distribution} (\cref{fig:semisup_feature_distribution}).
Since reparameterization is applied in both the unsupervised and the semi-supervised encoder, the latent spaces are smooth and Gaussian-like in most dimensions.
The background estimate sampled by the generative model agrees well with the background events in the pseudo-data.
The signal model \WNHyyN{150} is included in both figures to illustrate the separation between signal and background in the latent space.
In \cref{fig:unsup_feat_3,fig:unsup_feat_5,fig:semisup_feat_3,fig:semisup_feat_5}, the signal distribution deviates strongly from the background distribution, creating the overdensity that the weakly supervised classifiers can exploit if such signal events are present in the data.

\begin{figure*}[htbp]
  \centering
  \subfloat[]{\includegraphics[width=0.32\textwidth,page=1]{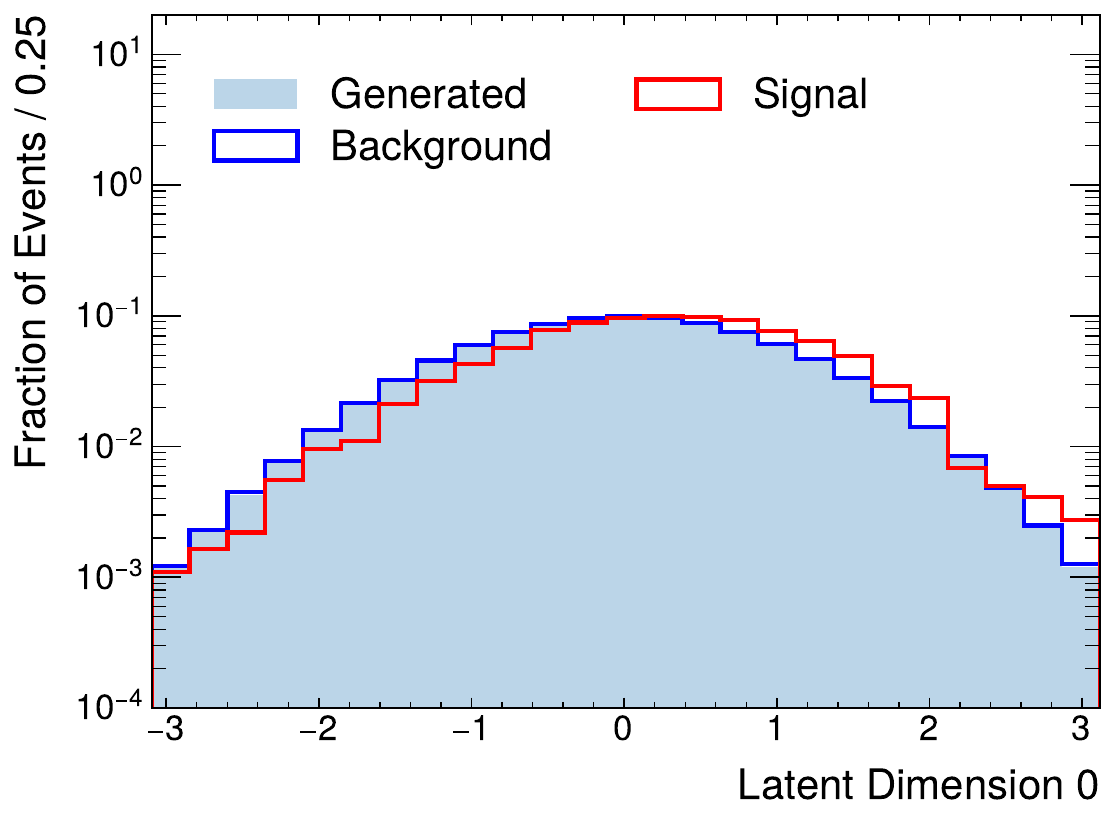}\label{fig:unsup_feat_1}}
  \hfill
  \subfloat[]{\includegraphics[width=0.32\textwidth,page=2]{figures/appendix_feature_dist/unsup_feature_distributions.pdf}\label{fig:unsup_feat_2}}
  \hfill
  \subfloat[]{\includegraphics[width=0.32\textwidth,page=3]{figures/appendix_feature_dist/unsup_feature_distributions.pdf}\label{fig:unsup_feat_3}}\\
  \subfloat[]{\includegraphics[width=0.32\textwidth,page=4]{figures/appendix_feature_dist/unsup_feature_distributions.pdf}\label{fig:unsup_feat_4}}
  \hfill
  \subfloat[]{\includegraphics[width=0.32\textwidth,page=5]{figures/appendix_feature_dist/unsup_feature_distributions.pdf}\label{fig:unsup_feat_5}}
  \hfill
  \subfloat[]{\includegraphics[width=0.32\textwidth,page=6]{figures/appendix_feature_dist/unsup_feature_distributions.pdf}\label{fig:unsup_feat_6}}\\
  \subfloat[]{\includegraphics[width=0.32\textwidth,page=7]{figures/appendix_feature_dist/unsup_feature_distributions.pdf}\label{fig:unsup_feat_7}}
  \caption{One-dimensional projections of the seven latent-space features for the unsupervised encoder. The distributions of the background from the pseudo-data (blue outline), the generated background (blue filled), and the \WNHyyN{150} signal (red) are compared. The background includes both the non-resonant and the resonant SM Higgs boson components.}
  \label{fig:unsup_feature_distribution}
\end{figure*}

\begin{figure*}[htbp]
  \centering
  \subfloat[]{\includegraphics[width=0.32\textwidth,page=1]{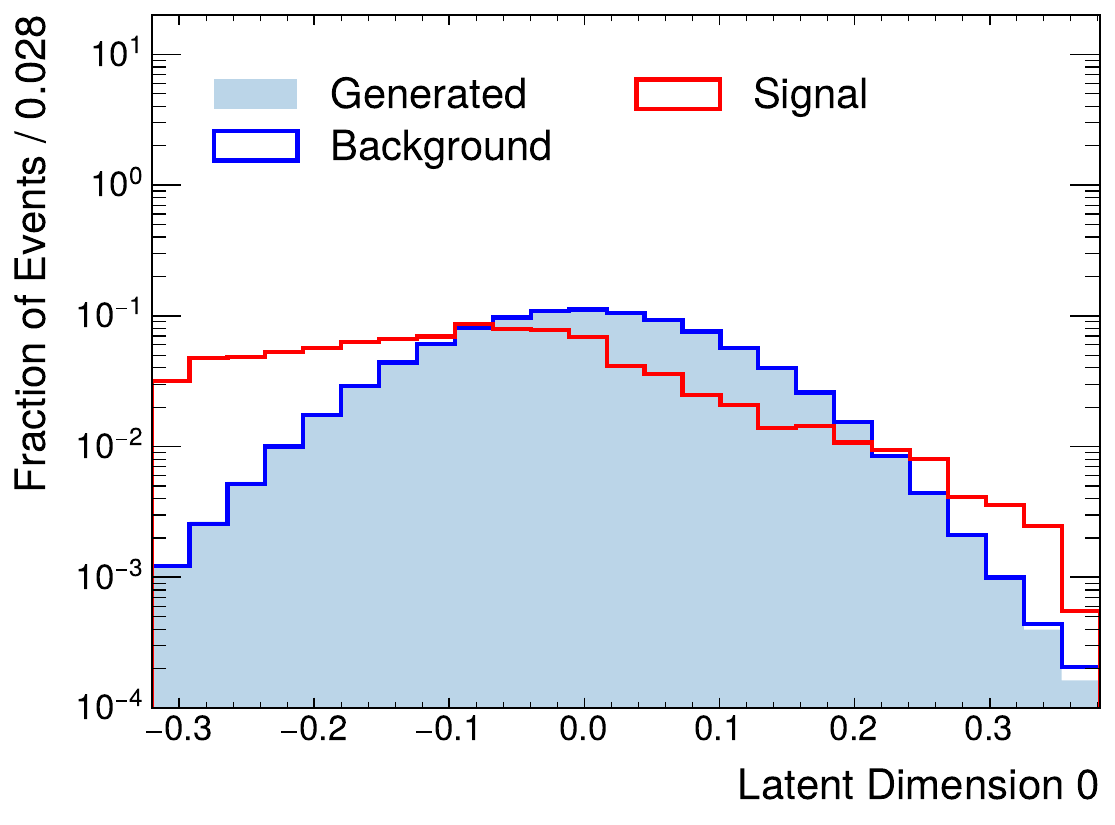}\label{fig:semisup_feat_1}}
  \hfill
  \subfloat[]{\includegraphics[width=0.32\textwidth,page=2]{figures/appendix_feature_dist/semisup_feature_distributions.pdf}\label{fig:semisup_feat_2}}
  \hfill
  \subfloat[]{\includegraphics[width=0.32\textwidth,page=3]{figures/appendix_feature_dist/semisup_feature_distributions.pdf}\label{fig:semisup_feat_3}}\\
  \subfloat[]{\includegraphics[width=0.32\textwidth,page=4]{figures/appendix_feature_dist/semisup_feature_distributions.pdf}\label{fig:semisup_feat_4}}
  \hfill
  \subfloat[]{\includegraphics[width=0.32\textwidth,page=5]{figures/appendix_feature_dist/semisup_feature_distributions.pdf}\label{fig:semisup_feat_5}}
  \hfill
  \subfloat[]{\includegraphics[width=0.32\textwidth,page=6]{figures/appendix_feature_dist/semisup_feature_distributions.pdf}\label{fig:semisup_feat_6}}
  \caption{One-dimensional projections of the six latent-space features for the semi-supervised encoder. The distributions of the background from the pseudo-data (blue outline), the generated background (blue filled), and the \WNHyyN{150} signal (red) are compared. The background includes both the non-resonant and the resonant SM Higgs boson components.}
  \label{fig:semisup_feature_distribution}
\end{figure*}

\section{Additional Cross Section Limits}
\label{app:limits_all}

\Cref{fig:limits_all} extends the summary in \cref{subsec:limits} and presents the model-dependent cross section upper limits for all signal models and mass parameters considered in this work.
As in \cref{fig:limits}, the limits obtained with the semi-supervised and the unsupervised embedding are shown together with the most sensitive cut-based selection for each signal.
All limits are derived with the signal-injection procedure described in \cref{sec:inference}.
\begin{figure}[tbh]
\centering
\includegraphics[width=0.95\textwidth,height=0.88\textheight,keepaspectratio]{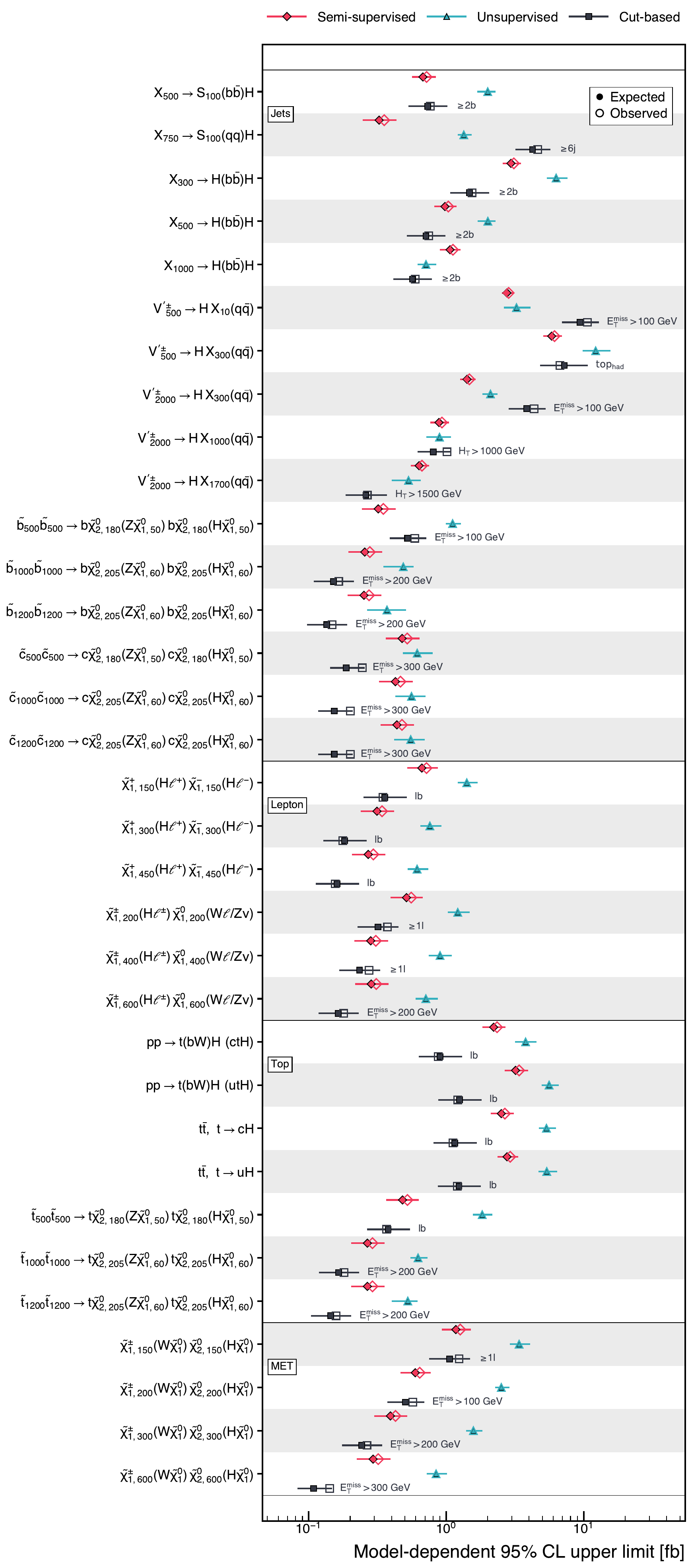}
\caption{Model-dependent cross section upper limits at 95\%~CL for all signal models and mass parameters considered in this work, comparing the semi-supervised and unsupervised embedding approaches to the most sensitive cut-based selection for each signal. Filled markers with their $\pm1\sigma$ bands show the expected limits and open markers the observed limits. Limits are projected to a total integrated luminosity of 470\,\ifb.}
\label{fig:limits_all}
\end{figure}

\section{Semi-supervised Encoder on Unseen Signals}
\label{app:semisupervised_encoder_extrapolation}

Since signal models are used in the training of the contrastive encoder to help it construct a latent space sensitive to new physics processes, a natural question is whether this encoder design remains sensitive to signal models not present in the training data.
Extrapolation is generally challenging in machine learning, and the performance typically depends on how far the extrapolation reaches.
For example, if the training set contains a signal model that closely resembles the extrapolation target, good performance on the target is more likely.
Such a study has been performed in Ref.~\cite{li2026signalawarecontrastivelatentspaces}, where individual signal models (either a single mass point or an entire signal process at all mass points) are removed from the training set, and the performance of the weakly supervised classifier is evaluated with the same \gls{CATHODE} pipeline.
When a single mass point of a signal model is excluded from the training set, the performance of the contrastive encoder is essentially unchanged, since the interpolation is well supported by the other mass points of the same process.
When an entire signal process is excluded at all mass parameters, the performance generally degrades more; however, for most of the signal models tested in Ref.~\cite{li2026signalawarecontrastivelatentspaces}, the discovery threshold can still be reached with an initial signal strength of less than 1$\sigma$.
More importantly, it has been shown that, compared to contrastive training with background processes only (which is genuinely signal-model agnostic), adding signal processes to the training generally improves the extrapolation.

To demonstrate the ability to generalize to unseen signals in the present work, we test the interpolation and extrapolation performance on two signal models.
For \cref{fig:interpolation_scan}, the signal model \WlZvHvHyyl{400} is removed from the training set of the contrastive encoder, while \WlZvHvHyylall~events at the other mass parameters, such as \WlZvHvHyyl{200} and \WlZvHvHyyl{600}, are retained. The full analysis pipeline is rerun to obtain the model-independent limit without signal injection in the pseudo-data, and \WlZvHvHyyl{400} is injected at different cross sections to obtain the model-dependent limit.
Comparing \cref{fig:scan_Wl_Hyyl400_intraining}, where \WlZvHvHyyl{400} is included in the training, with \cref{fig:scan_Wl_Hyyl400_holdout}, where it is held out, both the limit and the signal yield as a function of the injected cross section remain almost unchanged, showing that the semi-supervised encoder is robust for signals with mass parameters not included in the training sample.

\begin{figure*}[htbp]
  \centering
  \begin{subfigure}{0.48\textwidth}
    \centering
    \includegraphics[width=\linewidth]{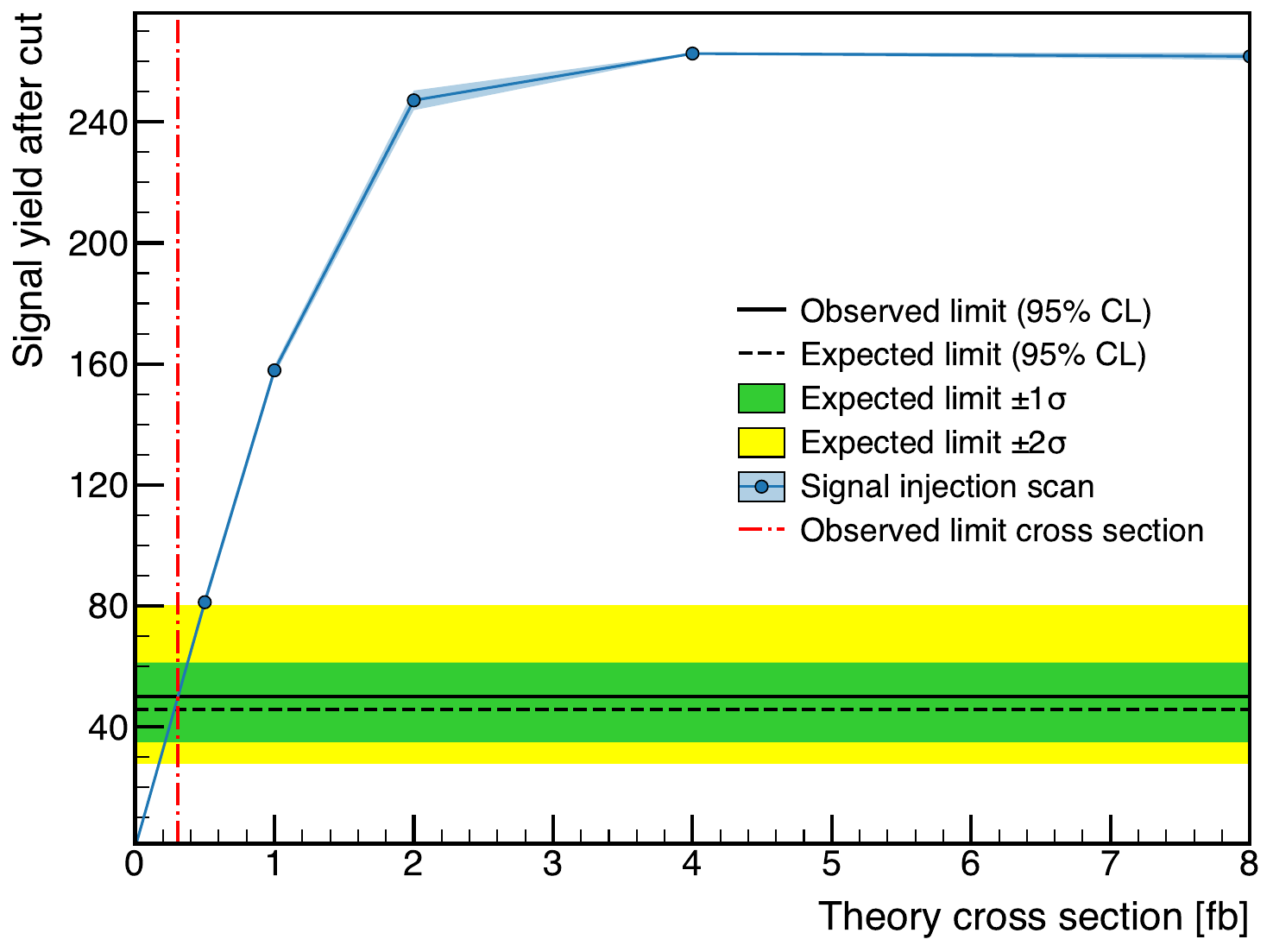}
    \caption{In Dataset}\label{fig:scan_Wl_Hyyl400_intraining}
  \end{subfigure}
  \hfill
  \begin{subfigure}{0.48\textwidth}
    \centering
    \includegraphics[width=\linewidth]{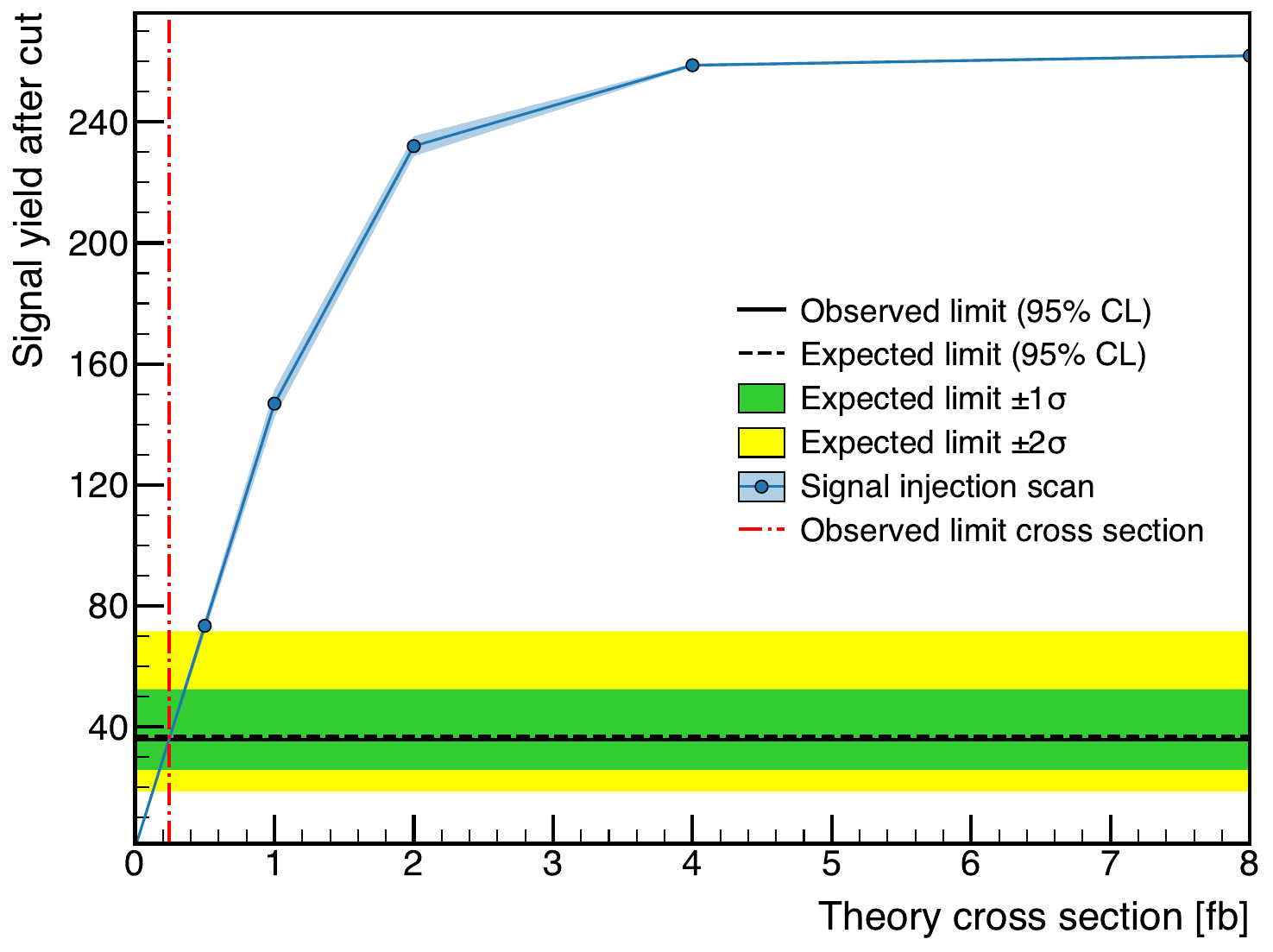}
    \caption{Interpolation}\label{fig:scan_Wl_Hyyl400_holdout}
  \end{subfigure}
  \caption{Limit-setting scan for \WlZvHvHyyl{400}. The number of injected signal events passing the selection is shown as a function of the injected cross section, together with the model-independent upper limit on the number of signal events $N$; the vertical dash-dotted line marks the resulting model-dependent cross section limit. \Cref{fig:scan_Wl_Hyyl400_intraining} shows the results with \WlZvHvHyyl{400} included in the training dataset of the contrastive encoder. \Cref{fig:scan_Wl_Hyyl400_holdout} shows the results with the \WlZvHvHyyl{400} mass point held out from the training.}
  \label{fig:interpolation_scan}
\end{figure*}

For the extrapolation test, shown in \cref{fig:extrapolation_scan}, all \HVTVcXjjHyyall~events are excluded from the training of the contrastive encoder, and the \HVTVcXjjHyy{2000}{300} model is tested, with the result shown in \cref{fig:scan_HVT_holdout}.
Compared to \cref{fig:scan_HVT_intraining}, where the \HVTVcXjjHyyall~process is included in the training, the signal yield as a function of the injected cross section drops, and the observed limit degrades from 1.47\,\ifb{} to 3.24\,\ifb{}.
For comparison, the most sensitive of the considered cut-based selections for this signal, $E_{\mathrm{T}}^{\mathrm{miss}} > 100\,\mathrm{GeV}$, results in an observed limit of 4.34\,\ifb{}, weaker than the limit obtained with the contrastive encoder extrapolated to this signal model.

\begin{figure*}[htbp]
  \centering
  \begin{subfigure}{0.48\textwidth}
    \centering
    \includegraphics[width=\linewidth]{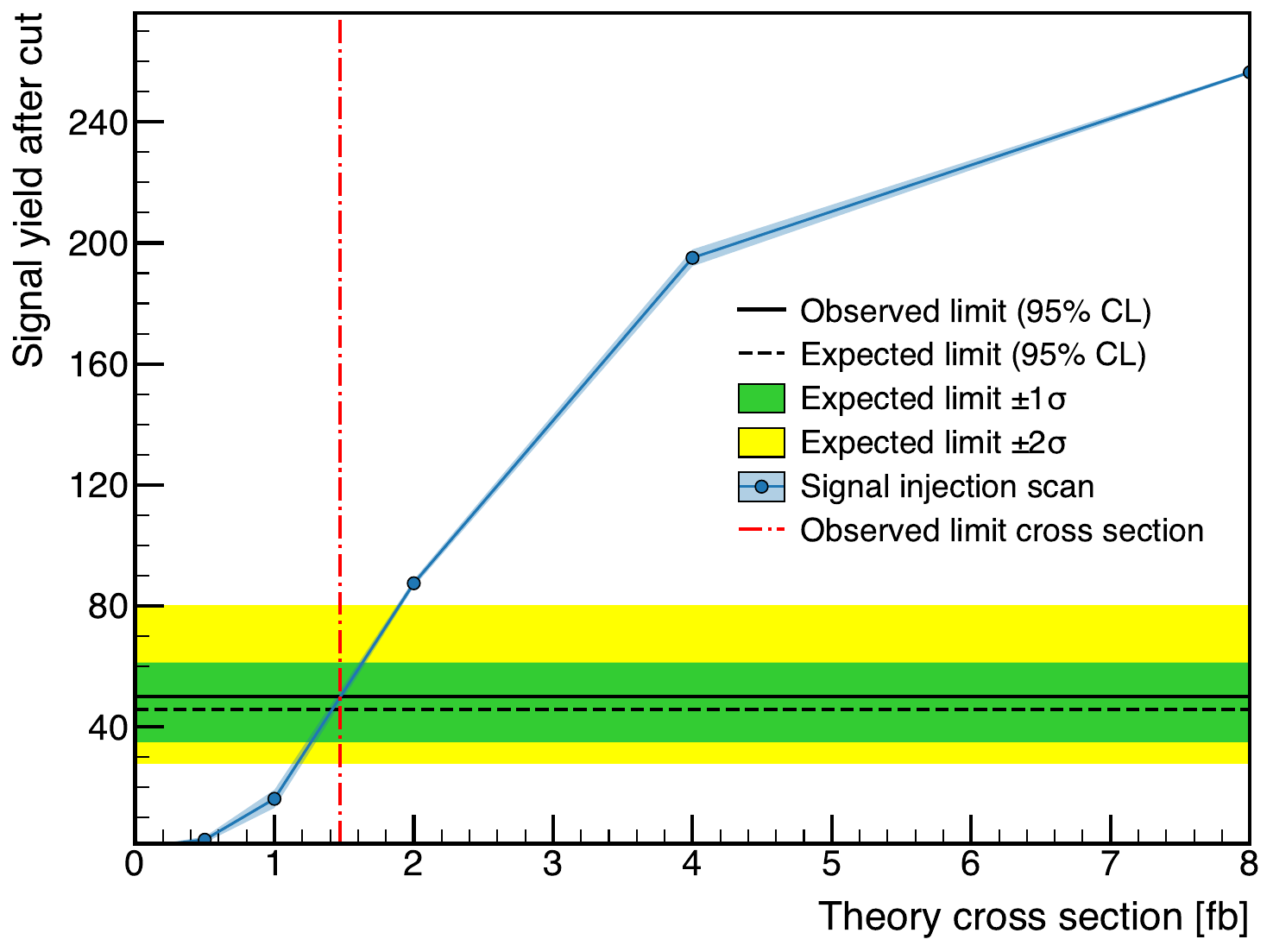}
    \caption{In Dataset}\label{fig:scan_HVT_intraining}
  \end{subfigure}
  \hfill
  \begin{subfigure}{0.48\textwidth}
    \centering
    \includegraphics[width=\linewidth]{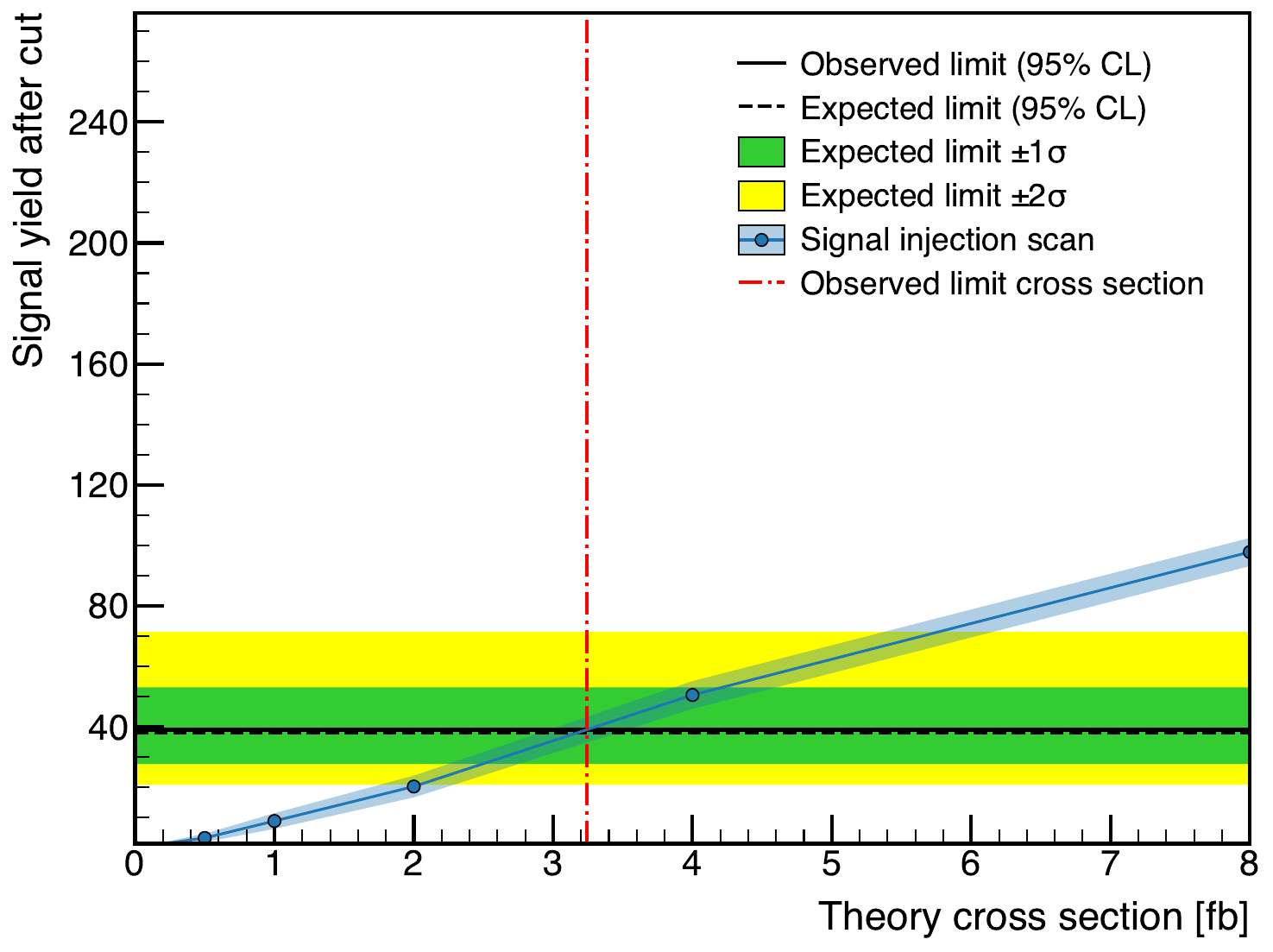}
    \caption{Extrapolation}\label{fig:scan_HVT_holdout}
  \end{subfigure}
  \caption{Limit-setting scan for \HVTVcXjjHyy{2000}{300}, presented as in \cref{fig:interpolation_scan}. \Cref{fig:scan_HVT_intraining} shows the results with \HVTVcXjjHyy{2000}{300} included in the training dataset of the contrastive encoder. \Cref{fig:scan_HVT_holdout} shows the results with all \HVTVcXjjHyyall~models held out from the training.}
  \label{fig:extrapolation_scan}
\end{figure*}

\end{document}